\documentclass[prx,aps,twocolumn,amsmath,footinbib,nobibnotes,superscriptaddress]{revtex4-2}
\usepackage{amssymb}
\usepackage{graphicx}
\usepackage{hyperref}

\usepackage{color} % for colored text

\definecolor{darkgreen}{rgb}{0,0.7,0}

\usepackage{booktabs}
\usepackage[table]{xcolor}
\definecolor{lightgray}{gray}{0.9}

\usepackage{soul} % for strikethrough text
\setstcolor{red}

\usepackage[caption=false]{subfig}
\newcommand{\phantomsubfloat}[1]{
    {% apply caption setup only temporarily
        \captionsetup[subfigure]{labelformat=empty}
        \subfloat[][]{#1}
    }%
}

\begin{document}

\title{Emergent Pauli blocking in a weakly interacting Bose gas}

\author{Federica Cataldini}
\thanks{These authors contributed equally to this work}
\affiliation{Vienna Center for Quantum Science and Technology (VCQ), Atominstitut, TU Wien, Vienna, Austria}
\author{Frederik M{\o}ller} 
\thanks{These authors contributed equally to this work}
\affiliation{Vienna Center for Quantum Science and Technology (VCQ), Atominstitut, TU Wien, Vienna, Austria}
\author{Mohammadamin Tajik}
\affiliation{Vienna Center for Quantum Science and Technology (VCQ), Atominstitut, TU Wien, Vienna, Austria}
\author{Jo\~ao Sabino}
\affiliation{Vienna Center for Quantum Science and Technology (VCQ), Atominstitut, TU Wien, Vienna, Austria}
\affiliation{Instituto Superior Técnico, Universidade de Lisboa, Lisbon, Portugal}
\affiliation{Instituto de Telecomunicações, Physics of Information and Quantum Technologies Group, Lisbon, Portugal}
\author{Si-Cong~Ji}
\affiliation{Vienna Center for Quantum Science and Technology (VCQ), Atominstitut, TU Wien, Vienna, Austria}
\author{Igor Mazets}
\affiliation{Vienna Center for Quantum Science and Technology (VCQ), Atominstitut, TU Wien, Vienna, Austria}
\affiliation{Research Platform MMM "Mathematics---Magnetism---Materials", \\ c/o Fakult\"at f\"ur Mathematik, Universit\"at Wien, 1090 Vienna, Austria} 
\affiliation{Wolfgang Pauli Institut,  c/o Fakult\"at f\"ur Mathematik, Universit\"at Wien, 1090 Vienna, Austria}
\author{Thomas Schweigler}
\affiliation{Vienna Center for Quantum Science and Technology (VCQ), Atominstitut, TU Wien, Vienna, Austria}
\affiliation{JILA, University of Colorado, Boulder, Colorado, USA}
\author{Bernhard Rauer}
\affiliation{Vienna Center for Quantum Science and Technology (VCQ), Atominstitut, TU Wien, Vienna, Austria}
\affiliation{Laboratoire Kastler Brossel, ENS-Universit\'e PSL, CNRS, Sorbonne Universit\'e, Coll\`ege de France, Paris, France}
\author{J\"{o}rg Schmiedmayer}
\affiliation{Vienna Center for Quantum Science and Technology (VCQ), Atominstitut, TU Wien, Vienna, Austria}

\begin{abstract}
The relationship between many-body interactions and dimensionality is integral to numerous emergent quantum phenomena.
A striking example is the Bose gas, which upon confinement to one dimension (1D) obeys an infinite set of conservation laws, prohibiting thermalization and constraining dynamics.
In our experiment, we demonstrate that such 1D behavior can extend much farther into the dimensional crossover towards 3D than expected.
Starting from a weakly interacting Bose gas trapped in a highly elongated potential, we perform a quench to instigate dynamics of a single density mode.
Employing the theory of Generalized Hydrodynamics, we identify the dominant relaxation mechanism as the 1D dephasing of the relevant collective excitations of the system, the rapidities.
Surprisingly, the dephasing remains dominant even for temperatures far exceeding conventional limits of one-dimensionality where thermalization should occur. 
We attribute our observations to an emergent Pauli blocking of transverse excitations, caused by the rapidities assuming fermionic statistics, despite the gas being purely bosonic.
Thus, our study suggests that 1D physics is less fragile than previously thought, as it can persist even in the presence of significant perturbations.
More broadly, by employing the exact Bethe ansatz solutions of the many-body system, we facilitate an interpretation of how the emergent macroscopic behavior arises from the microscopic interactions.

\end{abstract}

\maketitle 

\section{Introduction}

One-dimensional (1D) integrable systems offer a unique platform for studying many-body phenomena, as several of their properties can be computed exactly via the Bethe ansatz~\cite{Bethe1931}.
Its solutions are parameterized in terms of quasi-momenta, the rapidities $\theta$, encoding an extended set of conserved quantities.
Physically, the rapidities are the asymptotic momenta of microscopic scattering processes, whereby they collectively depend on interactions among many particles~\cite{PhysRevLett.80.3678, PhysRevLett.114.125302}. 
Further, to satisfy boundary conditions of the many-body wavefunction, the rapidities can not assume the same value, effectively making them obey fermionic statistics~\cite{lieb1963exact, doi:10.1063/1.1664947}. 
Such a fermionization at arbitrary interaction strength between atoms is a manifestation of the restrictions stemming from the the 1D geometry; in 3D, fermionization occurs only in the unitary limit~\cite{PhysRevLett.88.210403}.
Thus, the rapidity statistics, and the consequences thereof, are emergent properties of the system.
The evolution of the rapidities determines the integrable many-body dynamics, resulting in a description where conservation laws play a ubiquitous role~\cite{Rigol2008, Gogolin2016, Gring1318}.

Many-body systems in reduced dimensions can be realized experimentally through ultracold atomic gases~\cite{RevModPhys.80.885}.
A Bose gas in a tight transverse confinement, whose level spacing far exceeds all internal energy scales of the gas, will be restricted to the transverse ground state, effectively realizing a 1D integrable system~\cite{PhysRevLett.87.130402, PhysRevLett.87.160405, doi:10.1126/science.1100700, PhysRevLett.105.265302}.
As these energy scales approach the transverse level spacing, excitations in the transverse confinement become energetically possible.
If the behavior of the whole system remains 1D in character, whereby the transverse excitations can be treated perturbatively, the system is regarded as \textit{quasi}-1D~\cite{Gerbier_2004}.
Once such perturbative treatments break down, we consider the gas as truly three-dimensional~\cite{PhysRevA.65.043614}.
This occurs, for instance, when the transverse excitations exhibit collective behaviour, thereby requiring a description through 3D Bogoliubov theory.

Unlike integrable dynamics, the transverse excitations are not rapidity conserving and can thus lead to thermalization~\cite{10.21468/SciPostPhys.9.4.058}.
Recent kinematic approaches offer a description of integrability-breaking scattering processes consistent with Fermi’s golden rule~\cite{PhysRevLett.127.130601, bastianello2021hydrodynamics, PhysRevB.101.180302, PhysRevX.9.021027}.
The description permits a particular scenario owing to the fermionic nature of the occupied rapidities: If one of the rapidities of an outgoing scattering state is already occupied, the process becomes Pauli blocked.
This mechanism would allow integrable dynamics to persist at much longer time scales, thus enabling one-dimensionality to extend far beyond conventional energy scales~\cite{PhysRevA.83.021605}.
Importantly, the mechanism is entirely different from hard-core bosons~\cite{Paredes2004} and dynamical fermionization~\cite{Wilson2020}, as it relies solely on the quantum statistics of the rapidities occupation numbers present even in weakly interacting Bose gases.

\begin{figure}
\center
\includegraphics[width = \columnwidth]{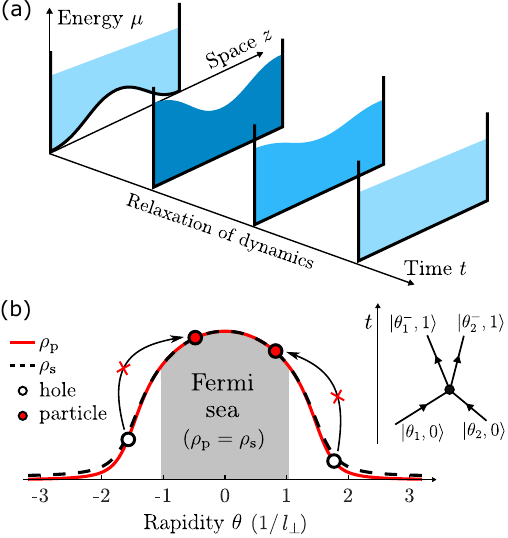}
\phantomsubfloat{\label{fig:quench_illustration_a}}
\phantomsubfloat{\label{fig:quench_illustration_b}}
\vspace{-2\baselineskip}% Remove extra line inserted by subfloat

\caption{\label{fig:quench_illustration}
(\textbf{a}) Illustration of the geometric quench setup. The gas is initialized in a 1D box trap whose bottom is sinusoidal. At time $t=0$ the potential is quenched to a flat box, initiating the dynamics.
(\textbf{b}) Schematic of emergent Pauli blocking.
Excitations in the transverse confining potential can occur following collisions between quasi-particles with large, opposite rapidities $| \theta_2 - \theta_1| \geq \sqrt{8}/l_\perp$. Here an excitation from the ground state $| 0 \rangle$ to the first excited state $| 1 \rangle$ is depicted.
The process correspond to creating two particle-hole pairs with rapidities $\theta_{1,2}^{-}$ and $\theta_{1,2}$.
If the rapidities of the outgoing particles are already occupied, the transverse excitation can not occur by virtue of the fermionic quasi-particle statistics.}
\end{figure}

To experimentally test how the emergent fermionic statistics may influence the physical behavior of the system, we realize a quasi-1D, weakly repulsively interacting Bose gas in a box trap.
The chemical potential and the thermal energy can be tuned to the order of, or exceeding, the transverse level spacing.
By controlling the shape of the bottom of the box trap, we can imprint a density perturbation in the form of a single cosine mode. Following a sudden quench to a flat box, the imprinted perturbation evolves and eventually relaxes, as illustrated in Fig.~\ref{fig:quench_illustration_a}.
Reducing the dynamics to the evolution of a single density mode drastically simplifies the study of the ensuing relaxation, thus making the setup an excellent probe for integrability breaking effects.

\section{Theoretical modeling}

While the dynamics of the gas following the quench is immensely complex on the microscopic level, on large scales it exhibits emergent hydrodynamic behavior described by the recent theory of Generalized Hydrodynamics (GHD)~\cite{castro2016emergent, bertini2016transport}.
Starting from equilibrium, the thermodynamic Bethe ansatz encodes the thermodynamic properties of a local equilibrium macrostate in a distribution of occupied rapidities $\rho_{\mathrm{p}} (\theta)$~\cite{doi:10.1063/1.1664947}.
Given $\rho_{\mathrm{p}} (\theta)$, thermodynamic expectation values of local operators can be computed, such as the atomic density $n = \int_{-\infty}^{\infty} \mathrm{d}\theta \: \rho_{\mathrm{p}}(\theta)$.
Occupied rapidities are considered fermionic quasi-particles with an infinite lifetime, while unoccupied rapidities are dubbed holes.
The sum of their respective densities yields the local density of states $\rho_{\mathrm{s}} (\theta)$.
Assuming local equilibrium, GHD provides a coarse-grained hydrodynamic description valid at large distance- and time-scales for systems near an integrability point.
Its validity has been demonstrated by several observations in 1D Bose gas experiments~\cite{schemmer2019generalized, malvania2020generalized, PhysRevLett.126.090602}.
According to the theory, the infinite set of continuity equations associated with the conserved charges can be written as a single kinematic equation for the quasi-particles 
\begin{equation}
    \partial_t \rho_{\mathrm{p}} + \partial_z (v^{\mathrm{eff}} \rho_{\mathrm{p}}) - \hbar ^{-1} \partial_\theta ( \partial_z U \: \rho_{\mathrm{p}}) = 0 \; .
\label{eq:GHD_main}
\end{equation}
Here, $U(z)$ is the external 1D potential, while the effective velocity $v^{\mathrm{eff}} (\theta)$ is the local propagation velocity of a quasi-particle with rapidity $\theta$. 
The effective velocity accounts for interactions between particles, which in integrable systems manifest as the Wigner delay time associated with the phase shifts occurring under elastic collisions~\cite{PhysRev.98.145, PhysRevB.97.045407}.
Hence, the emergent GHD of the 1D Bose gas comprises much richer physics than the conventional hydrodynamics~\cite{PhysRevLett.119.195301, bouchoule2022generalized}.
Expressions for GHD functions can be found in Appendix~\ref{app:theory}.

The integrable dynamics of GHD following Eq.~\eqref{eq:GHD_main} does not exhibit thermalization.
However, if the energy of an atomic collision exceeds twice the transverse level spacing, a transverse excitation, which breaks integrability, can occur. 
In our experiment the transverse trapping potential is parabolic and axially symmetric for the relevant excitation energies, whereby the coupling strength between two atoms is independent of their transverse state~\cite{Olshanii1998}. 
Therefore, the excitation of transverse states can be accounted for using a multi-component extension~\cite{Sutherland1968, klauser2011, doi:10.1142/S0217979202011895, PhysRevA.76.043617} of Yang’s theory~\cite{PhysRevLett.19.1312}; in addition to their rapidity, each quasi-particle carries a pseudo-spin encoding its transverse state~\cite{PhysRevLett.126.090602}.
Crucially, only a single rapidity distribution $\rho_{\mathrm{p}} (\theta)$ exists, which is shared for all transverse states.
Thus, two quasi-particles can not have the same rapidity even if they are in two different pseudo-spin states. 
Note that although only their rapidity component exhibits fermionic behaviour, we will be referring to the quasi-particles as fermionic in the following for brevity.
For integrability-breaking scattering processes one can associate the in- and out-states with particles and holes~\cite{PhysRevLett.127.130601}.
In the case of transversely exciting collisions, the process creates two particle-hole pairs, with the rapidities of the particles being much smaller than those of the holes, reflecting the gain in transverse potential energy.
Within GHD, this is accounted for by adding a Boltzmann-type collision integral $\mathcal{I} (\theta)$ to the right hand side of Eq.~\eqref{eq:GHD_main}.
The multi-component extension was originally derived and experimentally verified near the ideal Bose gas regime of the Lieb-Liniger model, however, it has also been shown to capture leading order processes in the quasi-condensate regime~\cite{PhysRevLett.126.090602}.

The aforementioned Pauli blocking can occur in systems either sufficiently close to the many-body ground state or, as in our case, in systems with a high chemical potential.
In either scenario all low rapidity states are filled, forming a Fermi sea~\cite{lieb1963exact}, leaving no holes available for the outgoing particles of the transverse excitations to occupy (see Fig.~\ref{fig:quench_illustration_b}).
In fact, the terms of the Boltzmann collision integral describing the excitation processes scale with the density of holes at the outgoing rapidities.
Thus, any occupation of low rapidities will lead to a suppression of the transverse excitations.
In a thermal state, higher rapidities become increasingly populated as temperature increases.
Hence, in the absence of the fermionic statistics, one would expect to observe relaxation of the system through thermalization, whose rate depends on temperature.

\section{Experimental setup}

%% Experimental sequence --

To achieve the scenario described above, we realize a quasi-1D gas of ultracold bosons ($^{87}\mathrm{Rb}$ atoms) on an atom chip~\cite{atomchips}. The chip produces a cigar-shaped magnetic trap featuring a tight transverse confinement with trapping frequency $\omega_\perp = 2 \pi \times 1.38 \, \mathrm{kHz}$ and width $l_\perp = \sqrt{\hbar / m \omega_\perp} = 0.29$~{\textmu}m.
The chemical potential of the gas is $\mu \approx (0.8 - 1.1) \, \hbar \omega_\perp$ while the interaction strength, characterized by the dimensionless Lieb-Liniger parameter $\gamma$~\cite{lieb1963exact}, is around 0.002, placing us fairly deep within the weakly interacting quasi-condensate regime.
The temperature of the gas can be adjusted by tuning the efficiency of the cooling and can be measured via the technique of density-ripples thermometry~\cite{PhysRevA.81.031610, PhysRevA.104.043305, https://doi.org/10.48550/arxiv.1908.00422}.
Heating and atom losses are negligible.

Using a digital micro-mirror device (DMD) we can create a desired optical dipole potential along the longitudinal axis of the trap~\cite{Tajik:19}.
We superpose two hard walls on the condensate, confining it to a region of $L = 80$~{\textmu}m. Between the walls we generate two different potentials: A flat potential and a cosine-shaped potential. Adjusting the amplitude of the cosine potential allows us to address the corresponding density mode at different strengths. 
By initializing the system in one configuration and then rapidly switching to the other we realize a geometric quench instigating the dynamics of the condensate (see Fig.~\ref{fig:quench_illustration}).
Following the quench, we measure the dynamical evolution of the density profile of the gas using absorption imaging after $2\,\mathrm{ms}$ of time of flight.
We denote the density profile averaged over multiple repetitions as $n(z,t)$.

\begin{figure*}
\center
\includegraphics[width = 0.98\textwidth]{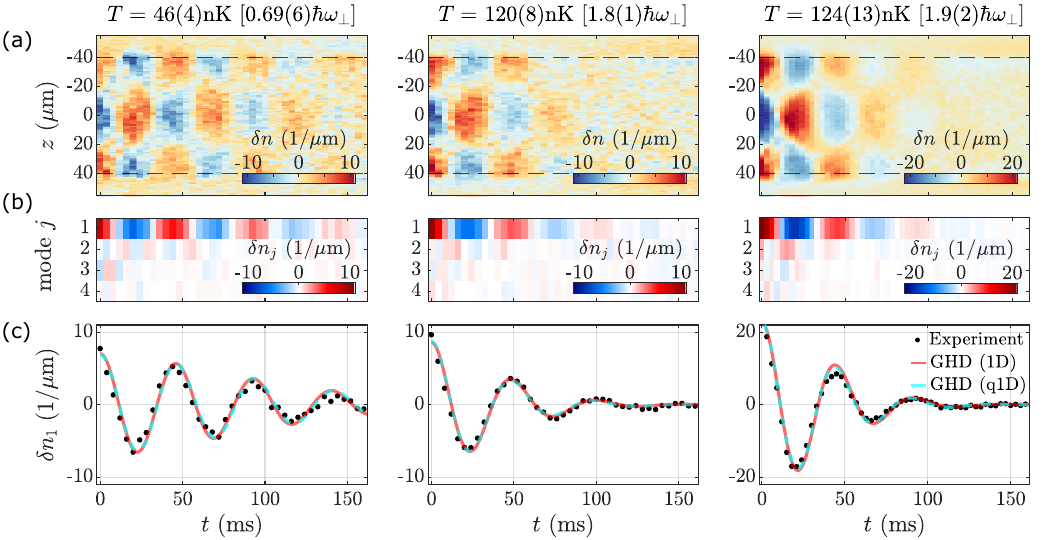}
\phantomsubfloat{\label{fig:carpets_and_modes_a}}
\phantomsubfloat{\label{fig:carpets_and_modes_b}}
\phantomsubfloat{\label{fig:carpets_and_modes_c}}
\vspace{-4\baselineskip}% Remove extra line inserted by subfloat

\caption{\label{fig:carpets_and_modes}
(\textbf{a}) Measured time evolution of the density perturbation $\delta n(z,t) = n(z,t) - \langle n(z,t)  \rangle_t$ for three separate quenches of the lowest cosine mode.
The mean atomic densities are 68, 60 and 82~{\textmu}m$^{-1}$, respectively.
The dashed lines mark the theoretical position of the hard walls. 
(\textbf{b}) Evolution of the lowest density modes $\delta n_j (t)$ obtained via a Fourier decomposition. 
(\textbf{c}) Evolution of the addressed mode $\delta n_1 (t)$ compared with both 1D and quasi-1D GHD simulations. 
}
\end{figure*}

\section{Dynamics of single density mode}

We perform three separate quenches, where we employ the DMD to imprint the lowest cosine mode of the box onto the condensate: (\textit{i}) Low temperature and small initial mode amplitude, (\textit{ii}) high temperature and small amplitude, and (\textit{iii}) high temperature and large initial mode amplitude~\footnote{For the two high temperature realizations, the bottom of the trap is switched from a cosine to a flat potential. For the low temperature realization, the order of trap configurations is reversed. For small quench amplitudes the evolution of the density perturbation only differs by a sign for the two quench types (see Fig.~\ref{fig:double_quench}). In Fig.~\ref{fig:carpets_and_modes} the sign for the low temperature realization is flipped for easier comparison.}.
For each quench the system is prepared in a thermal state~\cite{Hofferberth2008}, whose thermal energy scale, for the high temperature realizations, is close to twice the transverse energy gap.
Figure~\ref{fig:carpets_and_modes_a} shows the evolution of the density perturbation $\delta n(z,t) = n(z,t) - \langle n(z,t)  \rangle_t$ for each of the quenches, where $\langle \cdot \rangle_t$ denotes the average over time.
Following the quench at $t=0$ we find the evolution of each realization to exhibit damped oscillations in time, with the dynamics of those at higher temperature relaxing noticeably faster.

The density perturbation can be expressed as a sum of cosine modes $\delta n (z,t) = \sum_{j=0}^{\infty} \delta n_j(t) \cos \left( k_j z \right)$, where $\delta n_j (t)$ is the amplitude of the mode and $k_j = 2 \pi j /L$. 
Figure~\ref{fig:carpets_and_modes_b} shows the mode decomposition of the measured density perturbations for the three realizations. 
Only the four lowest modes are depicted, whose measurement is not affected by the finite resolution of our imaging apparatus.
In all three cases, the $j = 1$ density mode has a far greater population than any other mode, demonstrating that we indeed can address a single mode of the quasi-condensate with high accuracy.
The small population of the higher modes occur naturally in the thermal state, although small imperfections in the imprinted cosine potential will add to said population.

Figure~\ref{fig:carpets_and_modes_c} shows the evolution of the measured $j = 1$ density mode compared with GHD predictions~\cite{10.21468/SciPostPhys.8.3.041}.
The temperature and transverse trapping frequency (determining the coupling strength of the bosons) used in the simulations are obtained using separate measurements.
The theory is computed for a hard-walled box of length $L = 80$~{\textmu}m.
Due to the finite width of the experimentally realized walls, the measured density mode evolves in a box that is effectively a little longer.
We account for this by scaling the time axis of the simulations accordingly.
For more details on the simulations, see Appendix~\ref{app:exp_methods}.

Comparing the experimental observations to the 1D GHD predictions we observe a remarkably good agreement.
This is highly surprising, as the internal energy of the system (in particular for the two realizations with high temperature) are far beyond conventional conditions for one-dimensionality.
Under such conditions, thermalization through transverse excitations would normally have significant contribution to the dynamics.
However, when accounting for the Boltzmann-type collision integral in the calculations, we find that it hardly has any influence on the evolution of the density mode.
The reason is the vanishing density of holes in the Fermi sea of rapidities causing the excitation terms of the collision integral to vanish.
Note that de-excitations of thermally excited atoms in the initial state can still occur. 
For the results presented here, we assume no initial transverse excitations. 
Including an estimated thermal population of excited states only leads to slightly faster relaxation of the dynamics (see Fig.~\ref{fig:measurements}). 
Hence, the observed relaxation of the mode remains dominated by integrable processes, as the emergent Pauli blocking significantly prolongs the time-scale of thermalization.

\begin{figure}[b]
\center
\includegraphics[width = 0.98\columnwidth]{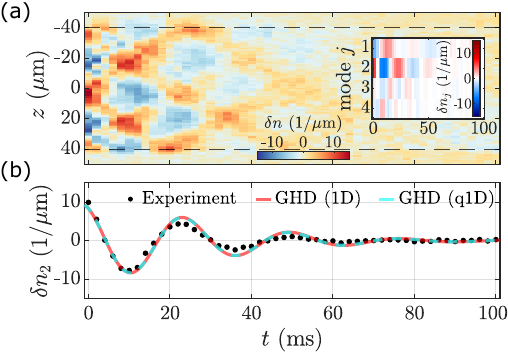}
\phantomsubfloat{\label{fig:carpets_modes_4th_single_a}}
\phantomsubfloat{\label{fig:carpets_modes_4th_single_b}}
\vspace{-2\baselineskip}% Remove extra line inserted by subfloat

\caption{\label{fig:carpets_modes_4th_single}
(\textbf{a}) Measured time evolution of the density perturbation $\delta n(z,t) = n(z,t) - \langle n(z,t)  \rangle_t$ for a quench of the $j=2$ cosine mode. The temperature is $T = 76(7)$~nK and the mean atomic density is 65~{\textmu}m$^{-1}$.
The dashed lines mark the theoretical position of the hard walls. 
The inset shows the evolution of the lowest density modes $\delta n_j (t)$. 
(\textbf{b}) Evolution of the addressed mode $\delta n_2 (t)$ with GHD theory comparison. 
}
\end{figure}

The observed behavior is not exclusive to the $j=1$ mode. Indeed, when exciting higher density modes (see Fig.~\ref{fig:carpets_modes_4th_single} for evolution of the $j=2$ mode) we find their dynamics to be captured by the 1D GHD as well~\footnote{The slightly slower relaxation of the GHD simulations compared to the experiment in Fig.~\ref{fig:carpets_modes_4th_single} is likely due to an underestimation of the temperature.}.
Additional sources of relaxation in the form of hydrodynamic diffusion could potentially become relevant for even higher modes~\cite{PhysRevLett.121.160603, PhysRevLett.125.240604}.
However, for the quenches explored here, we find their contribution negligible.

\section{Relaxation via dephasing of rapidities}

\begin{figure}
\center
\includegraphics[width = \columnwidth]{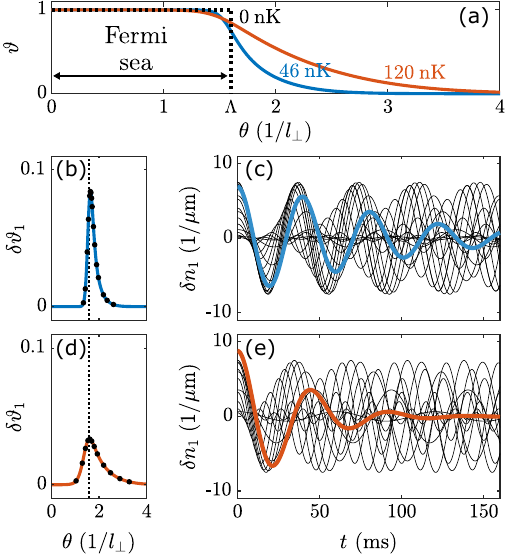}
\phantomsubfloat{\label{fig:dephasing_a}}
\phantomsubfloat{\label{fig:dephasing_b}}
\phantomsubfloat{\label{fig:dephasing_c}}
\phantomsubfloat{\label{fig:dephasing_d}}
\phantomsubfloat{\label{fig:dephasing_e}}
\vspace{-2\baselineskip}% Remove extra line inserted by subfloat

\caption{\label{fig:dephasing}
Relaxation of the excited density mode through dephasing of its rapidity components. 
(\textbf{a}) Occupation function $\vartheta (\theta)$ of a thermal state with density $n = 65\:${\textmu}m$^{-1}$ computed for the ground state ($T = 0\:$nK) and for temperatures corresponding to the small quench amplitude experimental realizations of Fig.~\ref{fig:carpets_and_modes}.
The occupation function is an even function of rapidity, therefore only positive rapidities are plotted here.
(\textbf{b, d}) $j=1$ cosine mode of the occupation function $\delta\vartheta_1 (\theta)$ of the low and high temperature realizations, respectively.
The dashed line indicates the Fermi momentum $\Lambda$ of the zero-temperature state.
The black dots mark select rapidities $\theta_i$, and the corresponding evolutions of $\delta\vartheta_1 (\theta_i)$ are plotted as thin, black lines in (\textbf{c, e}).
For reference, the simulated mode density $\delta n_1 (t)$ is plotted on top.
Note that in (c, e) the amplitudes of $\delta\vartheta_1 (\theta_i)$ are scaled to match $\delta n_1$.}
\end{figure}

Given that thermalization due to transverse excitations has a negligible influence at the time-scales explored here, the observed relaxation of the density mode must be a purely 1D phenomenon.
From the perspective of low-energy effective field theories, such as the Tomonaga-Luttinger liquid~\cite{PhysRevLett.47.1840, Haldane_1981} or Bogoliubov theory~\cite{PhysRevA.67.053615}, this may seem counter-intuitive, as the initial state of the experiment can be seen as a coherent population of a single eigenmode of the field theory Hamiltonian.
Within this picture, the excitations created by the experimental quench are long-wavelength phononic modes.
In the basis of phonon modes, which do not interact in the Luttinger liquid (or Bogoliubov) approximation, the excited eigenmode would not relax, see Appendix~\ref{app:LuttingerLiquid}.

\begin{figure*}
    \center
    \includegraphics[width = 1\textwidth]{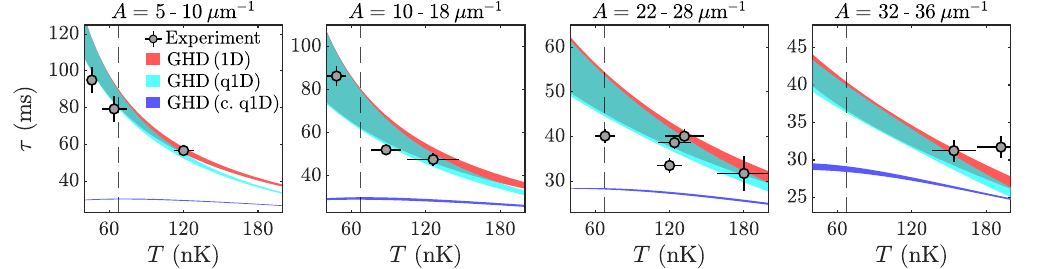}
    %\vspace{-2\baselineskip}% Remove extra line inserted by subfloat

    \caption{\label{fig:damping_scaling}
    Relaxation time-scale $\tau$ of the first density mode.
    For both experimental measurements (points) and GHD simulations (shaded areas), $\tau$ is obtained by fitting the time evolution of the $j=1$ density mode $\delta n_1 (t)$ with the damped oscillation of Eq.~(\ref{eq:damping_fit}).
    The experimental results are grouped into four ranges of mode amplitudes $A$. In each plot, $\tau$ is shown for different temperatures $T$. The temperature $T$ is inferred from density ripples analysis and the corresponding error bars represent the $68\%$ of confidence interval (more details can be found in Ref.~\cite{https://doi.org/10.48550/arxiv.1908.00422}). Meanwhile the error bars on the $\tau$ represent the $95\%$ of confidence interval from the fit to Eq.~(\ref{eq:damping_fit}).
    The dashed lines mark $k_B T = \hbar \omega_\perp$. 
    For each amplitude range, three different GHD simulations have been carried out: Standard 1D GHD (red), quasi-1D GHD accounting for the fermionic particle statistics (cyan), and quasi-1D GHD with a classical collision integral neglecting the statistics (blue).
    The GHD results are plotted as shaded areas, whose top and bottom edges mark the smallest and largest mode amplitude within the given range, respectively. 
    All simulations assume a mean atomic density of 75~{\textmu}m$^{-1}$, which deviates by up to $20\%$ for some of the measurements.
    Differences in mean density and effective box-lengths between simulation and experiment are accounted for by scaling the experimental time axis. 
}
\end{figure*}

To fully understand the nature of the apparent relaxation and why the phononic picture breaks down at the temperature scale of the experiment, consider the microscopic definition~\cite{korepin_bogoliubov_izergin_1993} of the sound velocity $v_{\mathrm{s}} \equiv \mathrm{lim}_{p \to 0} {\partial \varepsilon}/ {\partial p} $, where $\varepsilon (p)$ is the spectrum of elementary excitations with momentum $p$.
In a fermionic system near the ground state, low momentum excitations are found only in the vicinity of the Fermi momentum $\Lambda$.
According to the Bethe ansatz of the Lieb-Liniger model, the elementary excitations of the 1D Bose gas are the fermionic quasi-particles (rapidities)~\cite{LL2}, whereby the sound velocity reads~\cite{korepin_bogoliubov_izergin_1993}
\begin{equation}
    v_{\mathrm{s}}  ={  \frac{\partial_\theta \varepsilon (\theta)}{\partial_\theta p(\theta)}  \bigg\vert } _{\theta= \Lambda} \equiv v_{T = 0}^{\mathrm{eff}} (\Lambda) \; .
    \label{eq:sound_velocity}
\end{equation}
Hence, the sound velocity is equal to the effective velocity of GHD evaluated at the Fermi momentum for the many-body ground state.
As previously discussed, at zero temperature the quasi-particles fill up all the low-rapidity states, thus realizing a Fermi sea.
However, the edges of the Fermi sea start to melt as temperature increases, whereby the Fermi momentum becomes ill-defined. 
This is visualized in Fig.~\ref{fig:dephasing_a}, where the occupation function $\vartheta = \rho_{\mathrm{p}} / \rho_{\mathrm{s}}$ is plotted for the ground state of a Bose gas with density similar to the experiment.
The occupation function is defined as the occupied fraction of the allowed rapidity states, whereby $\vartheta = 1$ indicates a Fermi sea.
For comparison, occupation functions for temperatures matching the experimental realizations with low initial mode amplitudes shown in Fig.~\ref{fig:carpets_and_modes} are plotted.
In all three cases a Fermi sea can be found at low rapidities, however, for finite temperatures long thermal tails are present and no clear Fermi momentum can be identified. 
Therefore, at the temperature scales of the experiment, low-energy excitations can exist at a range of rapidities.
Such behaviour is not captured by the Luttinger liquid, where excitations are limited to the vicinity of the Fermi edge.

The effect of temperature (and initial mode amplitude) on the observed relaxation of the density mode can be illustrated as follows.
First, we compute the occupation functions $\vartheta (\theta, z)$ corresponding to the initial thermal states of the low initial mode amplitude realizations from Fig.~\ref{fig:carpets_and_modes}.
From their Fourier transforms, we extract the component of the occupation functions corresponding to the $j=1$ density mode $\vartheta_1 (\theta)$~\footnote{Assuming a linearization of the dressing operation around the stationary background (here, zeroth mode of the occupation function).}, which are plotted in Figs.~\ref{fig:dephasing_b} and \ref{fig:dephasing_d}.
Indeed, we find $\vartheta_1 (\theta)$ to have a significant width in rapidity space.
Thus, unlike the phononic modes of the effective free field theories, at finite temperature the excited density mode is carried by a distribution of quasi-momenta (rapidities)~\cite{Moller2022}.

The time-dependent solution of the occupation function for the perturbation can be obtained by linearizing the GHD equation around a stationary background~\cite{10.21468/SciPostPhysCore.1.1.002}.
For a single mode the solution reads (see Appendix~\ref{app:theory})
\begin{equation}
    \delta\vartheta_j (\theta, t) = \delta\vartheta_j (\theta, 0) \, e^{-i k_j  v_{0}^{\mathrm{eff}} (\theta)  t} \; ,
    \label{eq:momentum_mode_solution}
\end{equation}
where the effective velocity is evaluated using only the stationary background state.
The effective velocity is a monotonically increasing function of rapidity~\cite{LL2}, whereby each rapidity component of the mode will evolve at a slightly different rate.
Over time, this evolution will result in a gradual dephasing of the rapidity components, as shown in Figs.~\ref{fig:dephasing_c} and \ref{fig:dephasing_e}, which in turn leads to relaxation of the density mode.
Crucially, the occupation function of the perturbations is centered around the zero-temperature Fermi momentum, while their widths increase with temperature.
Hence, while the propagation velocity of the perturbation is (up to a small correction) given by Eq.~\eqref{eq:sound_velocity}, its relaxation is determined by the spread of involved rapidities and their dephasing according Eq.~\eqref{eq:momentum_mode_solution}. The higher the temperature, the larger this spread and, hence, the faster the relaxation. 
Indeed, this is exactly the behavior observed in the measured density carpets of Fig.~\ref{fig:carpets_and_modes_a}.
Thus, our analysis demonstrates how one can readily identify the various mechanisms of relaxation present within the system by employing the GHD and the Bethe ansatz.

\section{Time-scales of relaxation}

The dephasing longitudinal dynamics and the thermalizing transverse dynamics present two competing time-scales of relaxation in our system. 
To study the scaling of the two mechanisms, we perform a number of experimental quenches, and theoretical investigations thereof, for a wide range of temperatures and initial mode amplitudes.
We find that the observed dynamics of the $j=1$ density mode follows that of a damped oscillation.
In order to quantify the relaxation time-scale $\tau$, we fit the evolution of the density mode with the heuristic formula
\begin{equation}
    f (t) = A  \exp{\left[ - (t/\tau)^{3/2}\right]} \cos( \omega t + \phi ) \, .
    \label{eq:damping_fit}
\end{equation}
We have deduced empirically that the exponent $3/2$ produces a good fit to both the simulated and measured modes.
The results of our study are presented in Fig.~\ref{fig:damping_scaling}. For the integrable 1D GHD, we observe a faster relaxation for both higher temperatures and greater initial mode amplitudes, consistent with the greater spread in rapidity of the initial thermal state, see Fig.~\ref{fig:damping_scaling_2D}. 
Very similar relaxation rates are exhibited by the quasi-1D theory, indicating that thermalization occurs at time-scales much slower than dephasing in the system.
Only at temperatures three times larger than the transverse level spacing do we observe signs of transverse excitations. 
Between a total of thirteen different experimental quenches performed, we consistently observe agreement when comparing to either of the two theories. This is indicative of the highly controlled manner in which the state is prepared and quenched as well as an evidence of the robustness of the hydrodynamic description. A full overview of all the performed experiments and corresponding GHD simulations can be found in Fig.~\ref{fig:measurements}.

To demonstrate that the slow thermalization time-scale is indeed caused by the emergent Pauli blocking, we simulate the quasi-1D GHD employing a classical collision integral, which neglects the fermionic statistics of the quasi-particles. The resulting relaxation times presented in Fig.~\ref{fig:damping_scaling} are much faster and exhibit much weaker dependence on temperature and amplitude than both the quantum theories and the experiment.
Indeed, by virtue of the high chemical potential of the condensate, the system is already deep in the quasi-1D regime for even the weakest quenches and coldest temperatures realized. Hence, without the Pauli blocking mechanism, the dynamics relaxes through thermalization at a rate much faster than the dephasing of integrable 1D GHD.
This behavior is clearly not what we observe in the experiment, illustrating how the emergent behavior of the physical system is intrinsically linked to properties of the quasi-particles.
 
%------- Discussion.

Finally, it is important to realize that the observed prolonging of the thermalization time-scale is inherently related to the quench protocol employed: The box trap enables the Fermi sea to be established across the entire system.
Further, switching the potential shape mostly preserves the Fermi sea.
For contrast, in protocols like the quantum Newton's cradle only few of the low rapidities are occupied following the initial quench~\cite{kinoshita2006quantum, 10.21468/SciPostPhys.6.6.070}.
Indeed, following a cradle-like quench in another atom chip setup, clear signs of thermalization were observed~\cite{schemmer2019generalized, PhysRevLett.126.090602}.
Determining whether a Bose gas system is 1D is therefore not as straightforward as merely checking whether the
temperature and chemical potential fulfill $\mu, k_\mathrm{B} T < \hbar \omega_\perp$.
Ultimately, the effective dimensionality of these highly elongated systems is determined by the presence and nature of transverse excitations, which near the 1D limit must respect the quantum statistics of the rapidities.

\section{Conclusion}

To conclude, we have demonstrated that the integrable GHD accurately describes the dynamics of a Bose gas, whose chemical potential and thermal energy far exceed conventional limits for one-dimensionality.
While such a system would be expected to immediately thermalize, we instead observe a much slower relaxation of the dynamics consistent with a dephasing of the rapidity constituents of the excited mode. 
The rapidities themselves remain conserved due to an emergent Pauli blocking of the integrability-breaking scattering processes, here in the form of transverse excitations in the trap.
The fermionic nature of the rapidities, and thus the Pauli blocking, emerge as a consequence of the microscopic interactions.
Emergent quasi-particle descriptions are ubiquitous in quantum many-body physics.
Unique to integrable models, the quasi-particles provide an exact solution to the many-body system, facilitating a powerful interpretation of phenomena emergent from the complex microscopic details.
Thus, experimental studies of integrable systems in particular will enable a deeper understanding of the manifestation of emergent behavior.

\section*{Acknowledgments}
We thank S. Erne, I. Bouchoule, and J. Dubail for helpful discussions.

This research was funded in whole, or in part, by the Austrian Science Fund (FWF) and German Research Foundation (DFG) Research Unit FOR 2724 “Thermal machines in the thermal world.” For the purpose of open access, the author has applied a CC BY public copyright licence to any Author Accepted Manuscript version arising from this submission. The work was further supported by FQXI program on “Informations as fuel” ESQ Discovery Grant “Emergence of physical laws: from mathematical foundations to applications in many body physics” of the Austrian Academy of Sciences (\"{O}AW).
F.C., F.M., and J. Sabino acknowledge support by the Austrian Science Fund (FWF) in the framework of the Doctoral School on Complex Quantum Systems (CoQuS).
I.M. acknowledges the support by the Wiener Wissenschafts- und Technologiefonds (WWTF) via project No. MA16-066 (SEQUEX).  
S.-C.J. acknowledges support by an Erwin Schr\"{o}dinger Quantum Science and Technology (ESQ) Fellowship funded through the European Union’s Horizon 2020 research and innovation programme under Marie Sk\l{}odowska-Curie grant 801110.
J. Sabino acknowledges support by the Funda\c{c}\~ao para a Ci\^{e}ncia e a Tecnologia (PD/ BD/128641/2017).
T.S. acknowledges support from the Max Kade Foundation through a postdoctoral fellowship. B.R. acknowledges support by the European Union’s Horizon 2020 research and innovation programme under the Marie Sk\l{}odowska-Curie grant agreement No. 888707. \\

\appendix
\section{Experimental methods and data analysis} \label{app:exp_methods}

\subsection{Preparation of the initial state}
\begin{figure}
    \center
    \includegraphics[width = 1\columnwidth]{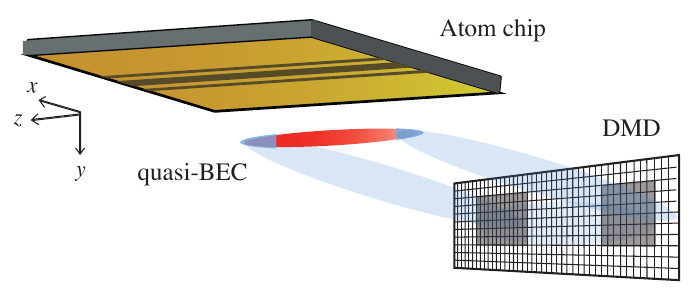}
    \caption{\label{fig:exp_setup} Illustration of the experimental setup. The cloud of $^{87}$Rb atoms is trapped below the atom chip whose current-carrying wires generate a cigar-shaped potential. In our convention, the longitudinal direction is labeled as $z$ and the transverse directions as $x$ and $y$. Here, gravity acts along the $y$-axis. A blue-detuned laser light is shined onto the atoms along the $x$ direction; using the DMD, its beam front is shaped to create the desired potential along the longitudinal axis.}
\end{figure}

\begin{figure}
    \centering
    \includegraphics[width = 1\columnwidth]{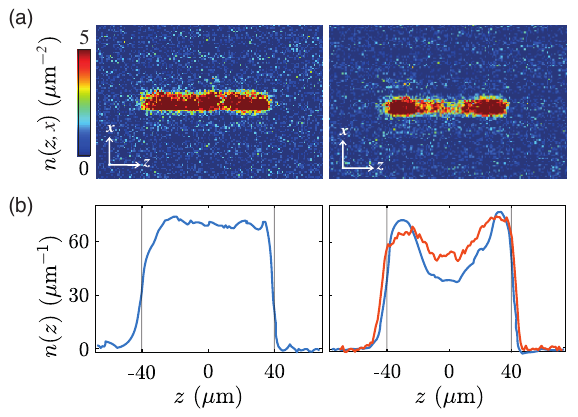}
    \caption{ \label{fig:density_profiles} \textbf{(a)} Absorption picture of the atomic cloud taken after 2~ms of TOF. The box length is $L=80$~{\textmu}m. The image shows the atomic cloud trapped in a flat-box potential (on the left), and in a cosine potential (on the right). The short TOF produces an expansion of the cloud in the transverse direction, due to the strong 1D confinement, while its effect on the longitudinal dimension is negligible. \textbf{(b)} Averaged longitudinal density profile for the cloud trapped in a flat box potential (on the left) and in a cosine potential (on the right). The two density profiles on the right side correspond to two different amplitudes.}
\end{figure}

We follow standard protocols of magneto-optical trapping, laser cooling and evaporative cooling to bring our system to the degeneracy temperature and realise a quasi-condensate of $^{87}$Rb atoms. The atom chip generates a RF-dressed cigar-shaped magnetic trap where the final stage of evaporative cooling is performed \cite{RauerThesis}. The magnetic potential has a fixed trapping frequency in the two tight (transverse) directions. 
To measure such frequency a sudden change in the current of the central chip wire (Fig.\ref{fig:exp_setup}) is performed, thus initiating an oscillation of the atoms in the magnetic trap.
%It can be measured by letting the atomic cloud oscillate in the trap after a sudden change in the current of the central chip wire (Fig.\ref{fig:exp_setup}). 
For all the measurements here presented $\omega_\perp = 2\pi \times 1.38 \, \mathrm{kHz}$.
On the other hand the geometry of the trapping potential in the elongated direction can be arbitrarily modified. To this purpose a blue-detuned laser light of frequency $660$~nm is overlapped to the magnetic trap along one of the two transverse directions. The shape of the dipole trap \cite{GRIMM200095}, and therefore of the effective longitudinal potential, can be designed by mean of a digital micro-mirror device (DMD) \cite{Tajik:19}. The setup implemented is schematically depicted in Fig.~\ref{fig:exp_setup}. In the presented experiments we used the DMD to imprint hard walls onto the atomic cloud and to shape the trap bottom either as a homogeneous potential of length $L = 80$~{\textmu}m, or as a sinusoidal modulation with a specific $k = 2\pi j/L $, with $j = 1,2$. In other words we can prepare the system either in a flat box potential or in a cosine potential (Fig.~\ref{fig:density_profiles}(b)). The amplitude of the sinusoidal modulation can be tuned with high accuracy such that the resulting density perturbation changes significantly. In our case the amplitude of the addressed mode varies between 10$\%$ and 40$\%$ of the mean homogeneous density. The observables needed to characterize the system are extracted through absorption imaging in time-of-flight (TOF). A short TOF of 2~ms is employed to extract the longitudinal density profile of the atomic cloud at each realization (see Fig.~\ref{fig:density_profiles}(a)). While to estimate the total number of atoms and the temperature of the gas we analyze the density ripples pattern emerging after 11.2~ms of TOF~\cite{PhysRevA.81.031610, PhysRevA.104.043305, https://doi.org/10.48550/arxiv.1908.00422}. The total number of atoms in the initial state can be tuned by adjusting the evaporative cooling radio-frequency.

\subsection{Geometric quench and dynamics}
\begin{figure}
    \center
    \includegraphics[width =0.88\columnwidth]{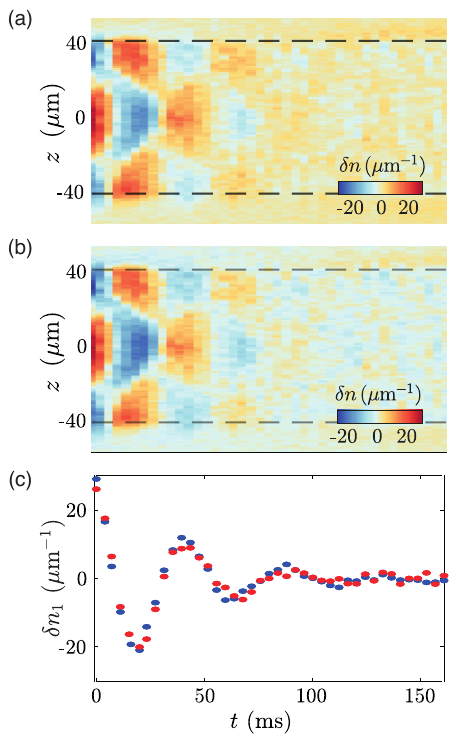}
    %\vspace{-0.5\baselineskip}
    \caption{\label{fig:double_quench} Symmetry of the geometric quench. The damping does not change if the dynamics is initialized in a cosine potential or in a flat box. In this measurement the mean atomic density is 97~{\textmu}m$^{-1}$ and the temperature of the initial thermal state is $T = 152$~nK. In (\textbf{a}) is illustrated the evolution of the density perturbation $\delta n(z,t)$ after a quench from the cosine-mode potential to a flat box; in (\textbf{b}) the quench is reversed. Figure (\textbf{c}) shows the evolution of the excited mode $\delta n_1(t)$. The blue data points correspond to the quench from single mode to flat box, while the red ones correspond to the reversed one (with flipped sign). Theoretically the symmetry of the two quenches is easily understood since the GHD model is invariant under time-reversal transformation.}
\end{figure}

\begin{figure}
    \centering
    \includegraphics[width =1\columnwidth]{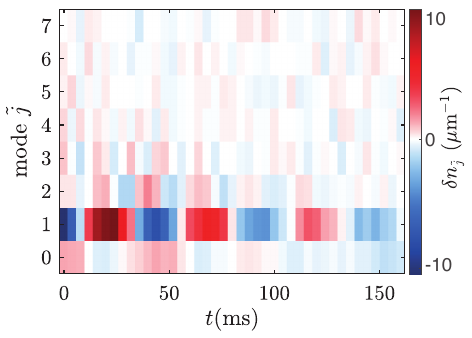}
    \caption{\label{fig:FFT_full} Fourier spectrum of the measured density perturbation of coldest quench illustrated in Fig.~\ref{fig:carpets_and_modes}. In the GHD simulations the odd modes vanish, justifying the expansion of the perturbation as a sum of even modes.}
    \label{fig:my_label}
\end{figure}

%The dynamics is initiated by imprinting a density perturbation to the system through a geometric quench.
For every experiment we design two different optical dipole potentials, a cosine shaped potential and a flat box. We initialize the dynamics by preparing the system in one of the two configuration and subsequently quenching to the other. Within our regime of temperatures and amplitudes, the order in which the quench is performed does not influence the emerging dynamics and the damping rate, as it is shown experimentally in Fig.~\ref{fig:double_quench}. The switch between the two potentials occurs in a few microseconds which is much faster than the timescale of the longitudinal dynamics of the system.
To trace the evolution of the perturbation over time we extract the longitudinal density profile of the quasi-condensate at each time step and consider its averaged value after several experimental repetitions. Since we are limited by shot noise we need a statistical ensemble of around 100 realizations to measure the expectation value. On average the variation of the atom number is about 15$\%$ of the mean value. \\
Given the evolution of the mean atomic density $n(z,t)$ (measured or simulated), we compute the density perturbation $\delta n(z,t) = n(z,t) - \langle n(z,t)  \rangle_t$, where $\langle \cdot \rangle_t$ denotes the average over time. For long enough evolution times $\langle n(z,t)  \rangle_t$ is equal to the asymptotic density profile. To extract the amplitudes of the cosine mode we take the Fourier transform of $\delta n(z,t) $ within the central $80$~{\textmu}m of the system for every time $t$ and compute the single-sided amplitude spectrum. Examining the Fourier spectrum we find the amplitudes of the odd modes to be very small (see Fig.~\ref{fig:FFT_full}). Their population is due to small asymmetries in the longitudinal box potential.
During the 160 ms of evolution the heating of the system is negligible and the measured atom loss rate is about 2 atoms/ms; it arises from three-body recombination, collisions with the background gas particles and technical noise. 

\subsection{Scaling of time axes}
We systematically find the oscillation frequencies of the experimentally measured density mode to be slightly lower than the GHD predictions. 
Examining the mode decompositions, we find the $j=0$ mode to oscillate out of phase with the $j=1$ mode, indicating that the dynamics extends past the theoretical position of the box-trap walls.
As seen in Fig.~\ref{fig:density_profiles}, the experimentally realized walls are not perfectly hard, but have a finite width due to experimental imperfections.
We assume that the finite wall width does not contribute to additional relaxation of the mode.
However, it does results in the density mode appearing to evolve in an effective box, whose length $L^{\mathrm{eff}}$ is slightly longer than the theoretical box $L = 80$~{\textmu}m. To accommodate for this, we can scale the experimental time axes by $L/L^{\mathrm{eff}}$.
For each quench, we estimate $L^{\mathrm{eff}}$ by fitting the mean density profile with two hyperbolic tangent functions. The wall width in the present measurements ranges from $2$~{\textmu}m to $5$~{\textmu}m each, which are then added to the theoretical box length $L$ in order to obtain $L^{\mathrm{eff}}$.

Further frequency differences between measurements and simulations can be found when the two feature a non-equal number of atoms. From the GHD perspective this is easily understood, as the effective velocity depends on the collective effects of interactions, which in turn depends on the local atomic density. To account for the difference, we find that the propagation velocity of the density perturbation (not to be confused with that of the quasi-particles) is very close to the speed of sound $v_{\mathrm{s}} = \sqrt{g n /m}$. Thus, when comparing the measured and simulated dynamics of realizations with different number of atoms, we scale the time axes of the measurements with the factor $\sqrt{\langle n_{\mathrm{sim}} \rangle_t / \langle n_{\mathrm{meas}} \rangle_t}$.

\begin{figure*}

    \center
    \includegraphics[width = 1\textwidth]{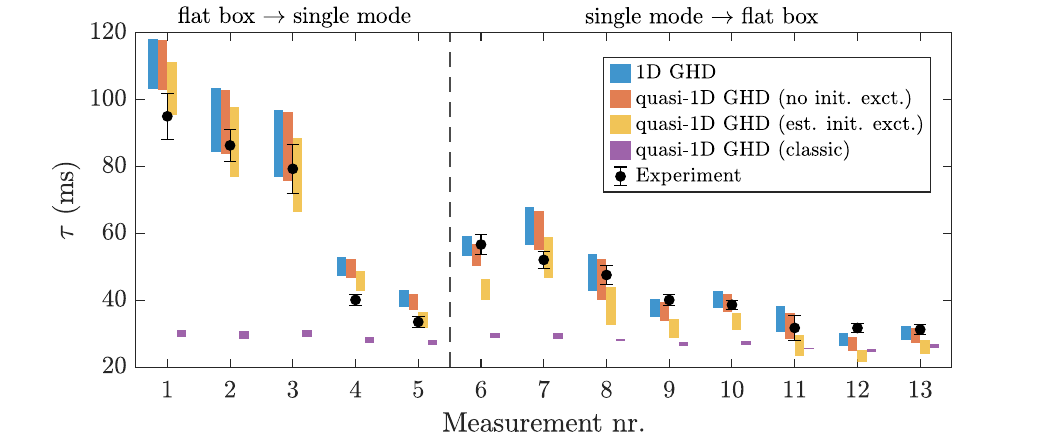}
    \vspace{-0.5\baselineskip}% Remove extra line inserted by subfloat
    
    \begin{minipage}{\textwidth}
    \begin{tabular}{llllllllllll}
    \toprule
    Nr.    & 1        & 2         & 3         & 4           & 5       & & &      \\ \midrule
    $T$ (nK) & 46(4)    & 48(8)     & 64(10)    & 68(8)       & 120(10)    & & &         \\
    $\langle n \rangle$ ({\textmu}m$^{-1}$)      & 68(2)     & 69(2)      & 62(2)      & 72(2)        & 83(2)   & & &           \\
    $A \;$ ({\textmu}m$^{-1})$      & 7.05(37) & 11.51(52) & 9.15(76)  & 25.04(1.10) & 25.69(1.35) & & &        \\
           &          &           &           &             &   & & &           \\ \midrule
    Nr.    & 6        & 7         & 8         & 9           & 10     & 11       & 12        & 13         \\ \midrule
    $T$ (nK) & 120(8)   & 88(12)    & 126(21)   & 132(16)     & 124(13)  & 180(22)     & 154(18)     & 192(19)    \\
    $\langle n \rangle$ ({\textmu}m$^{-1}$)      & 60(1)     & 74(1)      & 84(2)      & 72(1)        & 82(2)     & 74(1)        & 83(1)        & 77(1)    \\
    $A \;$ ({\textmu}m$^{-1})$      & 8.38(57) & 14.21(84) & 15.91(76) & 26.88(98)   & 24.87(97)   & 22.42(2.97) & 34.58(1.39) & 34.44(1.80)  \\
    \bottomrule
    \end{tabular}
    \end{minipage}

    % What happened??
    \caption{\label{fig:measurements}
    Relaxation time-scale of the first density mode. 
    For both experimental measurements (points) and GHD simulations (shaded areas), the time-scale $\tau$ is obtained by fitting the time evolution of the first density mode $\delta n_1 (t)$ with the damped oscillation of Eq.~\eqref{eq:damping_fit}.
    Four different GHD simulations have been carried out: Standard 1D GHD, quasi-1D GHD assuming no initial transverse excitations, quasi-1D GHD with an estimated thermal population of excited states, and quasi-1D GHD with a classical collision integral.
    The error bars of the experimental data points represent the $95\%$ confidence interval of $\tau$ from fitting with Eq.~\eqref{eq:damping_fit}.
    The extend of the shaded areas reflects the maximal variation in simulation results considering the uncertainties of both the mode amplitude $A$ and temperature $T$ listed in the table.
    The uncertainty of $A$ is given by the $95\%$ confidence interval from fitting with Eq.~\eqref{eq:damping_fit}. The measured temperature $T$ is inferred from density ripples analysis and the corresponding uncertainties represent the $68\%$ of confidence interval (more details can be found in Ref.~\cite{https://doi.org/10.48550/arxiv.1908.00422}).
    All simulations assume a mean atomic density of 75~{\textmu}m$^{-1}$ in the box.
    Meanwhile, the measured mean atomic densities $\langle n \rangle$, obtained by averaging over multiple experimental realizations, and their uncertainties given by the standard deviation can be found in the table.
    Differences in mean density and effective box-lengths between simulation and experiment are accounted for by scaling the experimental time axis. 
    }
\end{figure*}

\begin{figure}
    \center
    \includegraphics[width = \columnwidth]{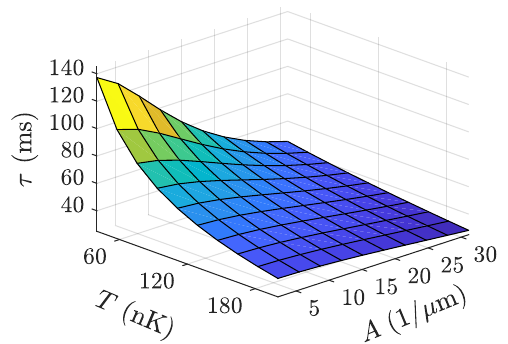}
    \vspace{0.5\baselineskip}
    \caption{\label{fig:damping_scaling_2D}
    Relaxation time-scale $\tau$ obtained by fitting the first density mode $\delta n_1 (t)$ from 1D GHD simulations with the damped oscillation of Eq.~\eqref{eq:damping_fit}. The fitted time-scale is shown as a function of temperature $T$ and initial mode amplitude $A$.
    By increasing either the temperature or the mode amplitude, the spread in rapidities of the density perturbation increases, thus resulting in a faster dephasing.
    }
\end{figure}

\section{Theoretical modelling of the system} \label{app:theory}

\subsection{Thermodynamic Bethe ansatz (TBA) of the Lieb-Liniger model}

An ultracold 1D gas of $N$ bosons with mass $m$ is described by the Lieb-Liniger Hamiltonian~\cite{lieb1963exact} plus an additional longitudinal potential $U(z)$  
\begin{equation}
    \mathcal{H}=-\frac{\hbar^{2}}{2 m} \sum_{i=1}^{N} \partial_{z_{i}}^{2}+g \sum_{i<j} \delta\left(z_{i}-z_{j}\right)+\sum_{i=1}^{N} U\left(z_{i}\right) \; ,
\end{equation}
where $g = 2 \hbar \omega_{\perp} a_s \left(1-1.03 a_s/l_{\perp} \right)^{-1}$ is the 1D contact interaction~\cite{Olshanii1998}, with $a_s$ being the 3D scattering length and $l_{\perp} = \sqrt{\hbar /(m \omega_\perp)}$.\\

Following the thermodynamic Bethe ansatz, for finite temperatures the local thermodynamic state of the Lieb-Liniger model can be fully encoded in a distribution of quasi-particles $\rho_{\mathrm{p}} (\theta)$, with each quasi-particle uniquely labeled by its rapidity $\theta$~\cite{doi:10.1063/1.1664947}.
Similarly, one can introduce a distribution of unoccupied rapidities, or holes, $\rho_{\mathrm{h}} (\theta)$ and the density of states $\rho_{\mathrm{s}} (\theta)$ obeying the relation
\begin{equation}
    \rho_{\mathrm{s}} (\theta) = \rho_{\mathrm{p}} (\theta) + \rho_{\mathrm{h}} (\theta) = \frac{1}{2 \pi} + \frac{1}{2 \pi} \int_{-\infty}^{\infty} \mathrm{d}\theta' \: \varphi(\theta - \theta') \rho_{\mathrm{p}} (\theta ') \; ,
\end{equation}
where $\varphi$ is the rapidity derivative of the two-body scattering phase given by
\begin{equation}
    \varphi(\theta-\theta')=\frac{2 m g / \hbar^2}{(m g / \hbar^2)^{2}+(\theta-\theta')^{2}} \; .
\end{equation}
Given the quasi-particle distribution, one can compute local thermodynamic expectation values of all conserved quantities of the system
\begin{equation}
    q_i = \int_{-\infty}^{\infty} \mathrm{d}\theta \: h_i(\theta) \rho_{\mathrm{p}} (\theta) \; ,
    \label{eq:conserved_charges}
\end{equation}
where $h_i$ is the single-particle eigenvalue of the $i$'th conserved quantity.
In our case, we are only interested in the expectation value of the atomic density, which is typically denoted as the $0$'th conserved quantity with $h_0 = 1$.

Further, one can introduce the occupation function $\vartheta (\theta) = \rho_{\mathrm{p}} (\theta) / \rho_{\mathrm{s}} (\theta)$ describing the fraction of allowed rapidities occupied. Since the Bethe ansatz quasi-particles of the Lieb-Liniger model obey fermionic statistics, a thermal state can be calculated following
\begin{equation}
    \vartheta (\theta) = \frac{1}{1 + e^{ \epsilon(\theta) \beta }} \; ,
\end{equation}
where $\beta = 1/k_B T$ is the inverse temperature and the pseudo-energy $\epsilon(\theta)$ is given by the relation
\begin{equation}
    \epsilon(\theta) = \frac{\hbar^2 \theta^2}{2 m} - \mu + \frac{1}{2 \pi \beta} \int_{-\infty}^{\infty} \mathrm{d}\theta' \: \varphi(\theta - \theta') \ln \left( 1 + e^{ \epsilon(\theta')  \beta } \right) \; .
\end{equation}
In the presence of an external potential $U(z)$ one can locally shift the chemical potential accordingly $\mu (z) = \mu_0 - U(z)$ under the local density approximation.

\subsection{Generalized Hydrodynamics (GHD)}

In GHD, the system is treated as a continuum of mesoscopic fluid cells in space-time, each in local equilibrium. The macroscopic flow between neighboring fluid cells occurs at a rate slower than the local microscopic relaxation, such that local thermodynamic equilibrium in the fluid cells is maintained at all times. Thus, each fluid cell is characterized by a local thermodynamic Bethe ansatz resulting in a time- and space-dependent quasi-particle distribution $\rho_{\mathrm{p}} (\theta, z, t)$~\cite{castro2016emergent,bertini2016transport}. Note, for brevity we omit all (equal) spacial and temporal arguments in the following.
The complicated dynamics of the interacting Bose gas can be solved using a single continuity equation \eqref{eq:GHD_main} for the analogous system of quasi-particles. 
A key component of this equation is the effective velocity
\begin{equation}
   v^{\mathrm{eff}}(\theta)= \frac{\hbar \theta}{m} + \int_{-\infty}^{\infty} d \theta^{\prime} \varphi(\theta - \theta')  \rho_{\mathrm{p}} (\theta') \left[v^{\mathrm{eff}}\left(\theta^{\prime}\right)-v^{\mathrm{eff}}(\theta)\right] \, ,
\end{equation}
representing the propagation velocity of a quasi-particle with rapidity $\theta$.
The propagation velocity encodes the Wigner delay time~\cite{PhysRev.98.145} associated with the quantum mechanical phase shifts occurring upon elastic collisions of the interacting atoms.
In the presence of integrability breaking mechanisms, such as transverse exctations, a Boltzmann-type collision integral can be added to the right hand side of Eq.~\eqref{eq:GHD_main}.

To illustrate how the relaxation of a single density mode in a hard-walled box-trap occurs via dephasing, we have employed a linearized version of GHD~\cite{10.21468/SciPostPhysCore.1.1.002}. 
First we split the time-dependent occupation function into a stationary background and an evolving perturbation $\vartheta (\theta, z, t) = \vartheta_0 (\theta) + \delta\vartheta(\theta, z, t)$.
The background can be identified as the zeroth mode of the occupation function, while the perturbation contains all higher modes.
If $\delta\vartheta \ll \vartheta_0$, we can neglect interactions during evolution within the perturbation itself and only treat the interactions between the perturbation and the stationary background. Thus, the GHD equation can be simplified as
\begin{equation}
    \partial_t \delta\vartheta (\theta, z, t) + v_{0}^{\mathrm{eff}}(\theta) \, \partial_z \delta\vartheta (\theta, z, t) = 0 \; .
    \label{eq:linearized_continuity}
\end{equation}
Here, the effective velocity is computed using only the background state, as signified by the subscript. Since the background state, and by extension the velocity field, is homogeneous, each Fourier mode evolves independently. Indeed, plugging a single mode $\delta\vartheta(\theta, z, t) = \delta\vartheta_j(\theta, t) e^{i k_j z}$ into Eq.~\eqref{eq:linearized_continuity} yields
\begin{equation}
    \partial_t \delta\vartheta_j (\theta, t) + i k_j v_{0}^{\mathrm{eff}} (\theta) \delta\vartheta_j (\theta, t) = 0 \; ,
\end{equation}
which has the time-dependent solution given in Eq.~\eqref{eq:momentum_mode_solution}.

\subsection{Transverse excitations as pseudo-spins}

Consider a transverse confining potential that is harmonic and axially symmetric for the relevant excitation energies.
This is, to a good approximation, the case of our experimental setup.
Since the anharmonicity of the potential is negligible, the center-of-motion degrees of freedom separate from the relative-motion ones. 
The relative motion of two atoms in the transverse plane is subject to the 2D Hamiltonian 
\begin{equation}
    \hat H_\mathrm{2D}=\frac {-\hbar ^2}{2m_\mathrm{r}}\left[ \frac 1{r_\perp }\frac {\partial ~}{\partial r_\perp }\left( r_\perp \frac {\partial ~}{\partial r_\perp }\right) +\frac 1{r_\perp ^2}\frac {\partial ^2}{\partial \varphi ^2}\right] + \frac {m_\mathrm{r}\omega _\perp ^2r_\perp ^2}2  , 
\end{equation}
where $m_\mathrm{r}=m/2$ is the reduced mass of the atomic pair, $r_\perp =\sqrt{x^2+y^2}$ is the 2D radius, and  $\varphi = \arctan (y/x)$ is the azimuthal angle. 

A collision due to contact interaction requires two atoms to come to the same spatial point; this is possible only if their relative-motion wave function is non-zero for $r_\perp =0$.
This is possible only when the principal quantum number $n$ for the relative motion is an even number and the angular momentum projection $\ell _z$ to the $z$-axis is zero.
The respective (angle-independent, normalized) wave function is~\cite{PhysRevA.79.024101} 
\begin{equation}
    \Phi _{n,0}(r_\perp )=\frac 1{\sqrt \pi l_\mathrm{r}}\exp \left( -\tfrac 12 r_\perp ^2/l_\mathrm{r}^2\right) L_n^{(0)}(r_\perp ^2/l_\mathrm{r}^2) \; , 
\end{equation}
where $L_n^{(0)}(R)\equiv L_n(R)$ is the Laguerre polynomial and $l_\mathrm{r}=\sqrt{\hbar /(m_\mathrm{r}\omega _\perp )}=\sqrt 2 l_\perp $. 
The interaction strength for the pair of atoms is proportional to $|\Phi _{n,0}(0)|^2$, $n=0,\, 2,\, 4,\, 6, \, \dots \, $. 
Since $L_n^{(0)}(0)=1$ \cite{mabramowitz64:handbook}, we have $|\Phi _{n,0}(0)|^2=1/(\pi l_\mathrm{r}^2)$ for all even $n$ (for our purpose the maximum transverse excitation energy is $2\hbar \omega_\perp $, whereby only $n=0$ and $n=2$ are relevant).
The independence of scattering properties of atoms on their transverse states in an axially symmetric, harmonic transverse confinement was used by Olshanii~\cite{Olshanii1998} in his analysis of the confinement induced resonance as a key feature allowing for obtaining analytic expression for $g$ for the arbitrary ratio $a_s/l_\mathrm{r}$.  
Therefore, under the conditions of our experiment, the coupling strength is the same for all pairs of colliding atoms, independently of their transverse states.

It is well-known (see, e.g., Ref.~\cite{PhysRevA.87.012707}) that if all the components of a multi-component bosonic system have the same mass and their two-body interaction is of the contact type (i.e., proportional to the $\delta $-function of the interparticle distance) with the strength independent of the component type of the colliding particles then the Yang-Baxter equation holds automatically.
The latter equation~\cite{jimbo1990yang} is a consistency condition that factorizes the three-body scattering matrix into two-body ones.
In our case, the independence of the interaction strength on the transverse states of the colliding pair of atoms is fulfilled accurately enough, and we can therefore treat our system as integrable.\\

A few words have to be said about the relation between transverse excitations and spin waves in sense of Ref.~\cite{PhysRevLett.19.1312}. 
Pairwise atomic collisions can change the symmetry type of the "coordinate" (longitudinal) and "spin" (transversal) parts of the bosonic $N$-body wave functions.
These symmetry types are described by irreducible representations (IR) of the symmetric group $S_N$ (the group of 
permutations of $N$ objects), uniquely denoted by Young diagrams~\cite{Bunker1998}.
The number of particles in each of the components of an integrable bosonic 1D multicomponent system are directly related to the lengths of different rows in the Young diagram denoting the particular IR, to which the spin function belongs.

Assume that initially all the atoms are in their radial ground state.
Their initial "spin" wave function $|0\rangle $ belongs to the fully symmetric IR with the Young diagram $\{ N \} $.
Next, assume that a collision of two fast atoms leads to an excitation of one atom to the radial level with the energy $2\hbar \omega _\perp $.
The final wave function is
\begin{equation}
    | \Psi_{\mathrm{f}} \rangle = \sqrt{ \frac {2}{N(N-1)}}\sum _{k>j}|\mathrm{S}\rangle _{jk}|\mathrm{Z}\rangle_{jk} \; , 
\end{equation}
where $|\mathrm{S}\rangle _{jk}$ and $|\mathrm{Z}\rangle _{jk}$ are, respectively, the spin (transverse) and coordinate ($z$-dependent) parts of the $N$-particle wave function after collision of the $j$th and $k$th atoms.
Since the sum is taken over all $N(N-1)/2$ pairs of atoms, the total wave function remains invariant against simultaneous permutation of both spins and $z$-coordinates of any pair of (bosonic) atoms.
Further, since both atoms in a pair experience a large momentum change after collision, the functions $|\mathrm{Z}\rangle _{jk}$ are mutually orthogonal for different $j,\, k$ pairs.
If one atom is brought to the second transversely excited state, the spin part is 
\begin{equation}
    |\mathrm{S}\rangle _{jk}=\frac 1{\sqrt 2} \Big(|2_j\rangle \prod_{l\neq j}| 0_l \rangle + |2_k\rangle \prod_{l\neq k}| 0_l \rangle \Big) \; , 
\end{equation}
where $|2_j\rangle \prod_{l\neq j}| 0_l \rangle$ denotes the $j$th atom in the second excited state and all the other atoms remaining in the ground state. 
Projecting $|\mathrm{S}\rangle _{jk}$ to the fully symmetric (i.e., corresponding to the Young diagram $\{ N \} $) transverse excited state $(1/\sqrt N)\sum _{j=1}^N |2_j\rangle $, we find a small overlap equal to $\sqrt{2/N}$. 
The rest of the $|\mathrm{S}\rangle _{jk}$ state corresponds to the Young diagram $\{ N-1,\, 1\} $. 
In a case of a collision that leads to excitation of both atoms to the first transverse excited level, the probability amplitude of the fully symmetric component in the final spin state is even smaller, being equal to $\sqrt{2 /[N(N-1)]}$.
We conclude therefore that a radial-state-changing collision leads with almost 100\% probability to the change of the Young diagram for the spin state, i.e., to the excitation of a "spin wave" in terms of Ref. \cite{PhysRevLett.19.1312}.

\subsection{Multi-component GHD and the collision integral}

In the quasi-1D regime, excited states of the transverse confinement can be populated through high energy collisions. We restrict our treatment to the three lowest states of the transverse potential and assume that a majority of atoms remains in the transverse ground state.
By virtue of parity, two possible excitation (and de-excitation) events are possible: (i) Two atoms in the ground state collide and both are excited to the first excited state, or (ii) two atoms in the ground state collide and one is excited to the second excited state. 
Thus, the total collision energy must exceed $2 \hbar \omega_\perp$ for transverse excitations to occur.

In the quasi-particle framework of integrable models, the collisions can be identified as scattering processes whose in- and out-states are given by holes and particles. In the context of a transverse excitation, the process can be understood as the creation of two particle-hole pairs, while the change in transverse state is treated as the change in psedo-spin state.
The $2 \hbar \omega_\perp$ gain in transverse potential energy following an excitation is reflected in the created particles having much lower rapidities compared to the holes, as the single-particle energy is given by $\varepsilon (\theta)= \hbar^2 \theta^2 /2 m$.

In response to the creation of holes or particles in an interacting integrable model, all local rapidities are shifted following the so-called backflow function~\cite{LL2}, modifying the associated collision integral~\cite{PhysRevLett.127.130601}.
However, in the ideal Bose gas phase of the Lieb-Liniger model, where the collision integral of Ref.~\cite{PhysRevLett.126.090602} originally was derived, the backflow is negligible.
In the quasi-condensate regime, contributions from the backflow may start becoming relevant, however, the "bare" (ideal Bose gas) collision integral should still represent leading order processes. Indeed, applying the "bare" collision integral to the experimental results of Ref.~\cite{schemmer2019generalized} (which featured a system comparable to ours) resulted in a better agreement with observations than purely 1D GHD~\cite{PhysRevLett.126.090602}. 

For the ideal Bose gas, the collision integral reads
\begin{equation} 
\begin{aligned}
    \mathcal{I}(\theta)= \sum_{n = 1}^{2} \frac{1}{2} \Big( &\left[ \mathcal{I}_{\mathrm{h}}^{+}(\theta)  -\mathcal{I}_{\mathrm{p}}^{-}(\theta) \right] + \\ 
    &\left[ \mathcal{I}_{\mathrm{h}}^{-}(\theta) -\mathcal{I}_{\mathrm{p}}^{+}(\theta)  \right] (\nu_{n})^{\beta_n}  \Big)
    \label{eq:collision_integral}
\end{aligned}
\end{equation}
where $\nu_n$ is the probability for an atom to be in the $n$'th transverse excited state, and $\beta_{1} = 2$ and $\beta_{2} = 1$ are the number of atoms changing state via the collisions.
The terms within the first set of square brackets of Eq.~\eqref{eq:collision_integral} describe the creation of quasi-particles and holes following transverse excitations. 
The effects of de-excitations are contained within the second set of square brackets. The terms are defined as
\begin{equation} 
\begin{aligned}
    \mathcal{I}_\alpha ^\pm (\theta)= \frac{(2\pi )^2\hbar}{m}\int _{\mathcal{R}_\pm }d\theta ^\prime \, \Big[
    \lvert\theta -\theta ^\prime \rvert P_\updownarrow (\lvert\theta -\theta ^\prime \rvert,\, \lvert\theta _\pm -\theta ^\prime _\pm \rvert) \\
    \rho_\alpha (\theta )\rho_\alpha (\theta ^\prime )\rho _{\bar \alpha}(\theta _\pm )\rho _{\bar \alpha }(\theta ^\prime _\pm ) \Big] \; ,
    \label{eq:collision_channel}
\end{aligned}
\end{equation} 
where $\bar \alpha  = {\mathrm{h}}$ for $\alpha = {\mathrm{p}}$ and vice versa, 
$P_\updownarrow (\theta _1,\, \theta _2)=4 c^2\theta _1\theta_2 /[\theta _1^2\theta_2^2 +c^2(\theta _1+\theta_2 )^2]$ is the scattering probability with $c=2a_s/l_\perp ^2$, while
$\theta _\pm = \frac 12 (\theta +\theta ^\prime )+\frac 12 (\theta -\theta ^\prime )
\sqrt{1\pm 8/[(\theta -\theta ^\prime )l_\perp ]^2}$ 
and $\theta _\pm ^\prime = \frac 12 (\theta +\theta ^\prime )-\frac 12 (\theta -\theta ^\prime )
\sqrt{1\pm 8/[(\theta -\theta ^\prime )l_\perp ]^2}$ are the rapidities after a collision leading to excitation  or de-excitation of the transverse states, respectively. 
The integration ranges in Eq.~\eqref{eq:collision_channel} are the following: 
$\mathcal{R}_+$ is the whole real axis, and $\mathcal{R}_-$ is comprised of those real values of $\theta ^\prime $, which yield real $\theta _- $ and $\theta _-^\prime $, i.e. 
$\mathcal{R}_- =\{ \theta ^\prime : \theta ^\prime < \theta -\sqrt 8/l_\perp \} \cup \{ \theta ^\prime : \theta ^\prime > \theta +\sqrt 8/l_\perp \}$.
Neglecting any heating effects, the excitation probabilities $\nu_{n} (t)$ follow the simple rate equations 
\begin{equation} 
\frac {d\nu_n }{dt}= \frac{1}{2} \beta_n \left[ \Gamma _{\mathrm{h}}^+ - \Gamma _{\mathrm{p}}^+ \nu_{n}^{\beta_n} \right], 
\label{nu12} 
\end{equation}
where $\Gamma _\alpha ^+=(2N)^{-1}\int _{-\infty }^\infty  dz \int _{-\infty}^\infty d\theta \, 
\mathcal{I}_\alpha ^+(\theta )$, $\alpha =\mathrm{p},\, \mathrm{h}$.

In the case where all allowed rapidities are occupied $\rho_{\mathrm{p}} (\theta) = \rho_{\mathrm{s}} (\theta)$, the density of holes vanishes for most incoming rapidities $\theta$ and $\theta'$ resulting in $\Gamma _{\mathrm{h}}^+ \approx 0$, meaning no transverse excitations can occur.
Meanwhile, for a non-degenerate 1D Bose gas, where the quasi-particle statistics become insignificant, we have $\rho _{\mathrm{p}}(\theta )\ll \rho _{\mathrm{h}}(\theta )\approx 1/(2\pi )$. Inserting this approximation into Eq.~\eqref{eq:collision_channel}, the collision integral in Eq.~\eqref{eq:collision_integral} takes the classical (Boltzmann) limit.

For high temperatures, additional transverse states beyond the three lowest may become relevant.
In this case, extending the multi-component model with additional components and collision channels may be possible, however, very cumbersome.
Meanwhile, if the product of the 1D atomic density and the 3D $s$-wave scattering length becomes larger than 1 then (i) the chemical potential begins to exceed twice the radial excitation energy and (ii) the effective 1D mean-field description is given by an essentially non-polynomial nonlinear equation.
In Ref.~\cite{PhysRevA.65.043614} an effective 1D mean-field equation is derived to describe the dynamics of a quasi-condensate under tight radial confinement in the high density regime.
The non-polynomial nature of this mean-field equation (replacing the 1D Gross-Pitaevskii equation) corresponds in the quantum regime to a Hamiltonian that substantially deviates from the Lieb-Liniger model and therefore precludes the use of the rapidity and other concepts inherent to 1D integrable theories.
In contrast, if the product of the 1D density to the scattering length is well below 1, the integrability breaking processes can be taken into account as a small perturbation of the integrable model~\cite{Mazets_2010}.
Clearly our setup corresponds to the latter case, as the evolution of the rapidity distribution (given by GHD) describes the observed dynamics well.

\section{Details of the numerical simulations}

All the calculations presented in the following were performed using the iFluid package~\cite{10.21468/SciPostPhys.8.3.041}.

\subsection{Fitting the initial state}
Following the evaporative cooling the system is well described by a thermal state~\cite{Hofferberth2008}, whose corresponding quasi-particle distribution $\rho_{\mathrm{p}} (\theta,z)$ can be computed using the thermodynamic Bethe ansatz~\cite{doi:10.1063/1.1664947}. The walls of the box potential are modelled as hard and infinitely tall and separated by a distance of $L = 80$~{\textmu}m. Further, for the potential between the walls a sinusoidal function is employed. We fix the number of atoms $N$, temperature $T$, and coupling constant $g$ according to their measured values. Then, we fit the amplitude of the sinusoidal potential and the chemical potential of the system to obtain the best match between the measured initial density profile and the theoretical one of Eq.~\eqref{eq:conserved_charges}.
The interaction strength is parameterized by the parameter $\gamma = m g / \hbar^2 n$, which for all realizations explored here is around $0.002$.
The thermal state is a non-linear function of the potential, meaning higher modes will also be populated initially, albeit much less than the addressed mode. We find good agreement between the measured and simulated dynamics of all modes, not just the one addressed by the potential.

\subsection{Time evolution}
At time $t=0$ we assume that the amplitude of the sinusoidal potential instantly becomes zero, thus realising a flat-bottomed box-trap. The subsequent evolution of the quasi-particle distribution is given by the GHD equation
\begin{equation}
    \partial_t \rho_{\mathrm{p}}  + \partial_z (v^{\mathrm{eff}} [\rho_{\mathrm{p}}] \, \rho_{\mathrm{p}}) = \mathcal{I}[\rho_{\mathrm{p}}] \; .  
    \label{eq:GHDpropagation_collisions}
\end{equation}
To numerically solve equation~\eqref{eq:GHDpropagation_collisions} we employ a split-step scheme: First, we propagate $\rho_{\mathrm{p}}$ a single time step $\Delta t$ (we used $\Delta t = 0.02\,\mathrm{ms}$) following $\partial_t \rho_{\mathrm{p}}  + \partial_z (v^{\mathrm{eff}} [\rho_{\mathrm{p}}] \, \rho_{\mathrm{p}}) = 0$ using the solution of characteristics. Next, we account for the transverse excitations by solving $\partial_t \rho_{\mathrm{p}} = \mathcal{I}[\rho_{\mathrm{p}}]$ for the same duration $\Delta t$.

For a box of length $L$, centered on $z = 0$, we model the hard walls by imposing the following boundary conditions
\begin{subequations}
\begin{align}
    \rho_{\mathrm{p}}(z = -L/2, \theta) &= \rho_{\mathrm{p}}(z = -L/2, -\theta) \label{eq:bound1} \\
    \rho_{\mathrm{p}}(z = L/2, \theta) &= \rho_{\mathrm{p}}(z = L/2, -\theta) \label{eq:bound2} \; ,
\end{align}
\end{subequations}
which can be interpreted as the quasi-particles having their rapidity reflected $\theta \to -\theta$ upon colliding with a wall.
For practical purposes we propagate the filling function $\vartheta$, rather than the quasi-particle density, for the first part of the split-step scheme.
To this end we employ the solution by characteristics, which reads
\begin{equation}
    \vartheta ( t', z, \theta  ) =  \vartheta ( t, \mathcal{U}(t',t, z, \theta), \mathcal{W}(t',t, z, \theta) ) \; ,
    \label{eq:propagation_by_characteristics}
\end{equation}
where the position and rapidity characteristics are given by 
\begin{align}
    \mathcal{U}(t',t, z, \theta) &= z - \int_{t}^{t'} \mathrm{d}\tau \, v_{\tau}^{\mathrm{eff}}\left( \mathcal{U}(\tau,t, z, \theta), \mathcal{W}(\tau,t, z, \theta) \right) \\
    \mathcal{W}(t',t, z, \theta) &= \theta - \int_{t}^{t'} \mathrm{d}\tau \, a_{\tau}^{\mathrm{eff}}\left( \mathcal{U}(\tau,t, z, \theta), \mathcal{W}(\tau,t, z, \theta) \right) \; ,
\end{align}
respectively. Here, the subscript $\tau$ denotes the dependence on the state at said time. Further, $a_{\tau}^{\mathrm{eff}} = -\hbar^{-1} \partial_z U(z)$ in our case. For the numerical simulation the we discretize the time axis in steps of $\Delta t$ and approximate the characteristics to first order~\cite{10.21468/SciPostPhys.8.3.041}.
To account for the boundary conditions~(\ref{eq:bound1},~\ref{eq:bound2}), the characteristics are modified in the following manner
\begin{equation}
  \mathcal{W}_{\mathrm{box}}(t', t, z,\theta) =
    \begin{cases}
      -\mathcal{W} & \text{for $\mathcal{U} - L/2 > 0$}\\
      -\mathcal{W} & \text{for $\mathcal{U} + L/2 < 0$}\\
      \mathcal{W} & \text{otherwise,}
    \end{cases}  
    \label{eq:charW}
\end{equation}
and 
\begin{equation}
  \mathcal{U}_{\mathrm{box}}(t', t, z,\theta) =
    \begin{cases}
      L - \mathcal{U} & \text{for $\mathcal{U} - L/2 > 0$}\\
      -L - \mathcal{U} & \text{for $\mathcal{U} + L/2 < 0$}\\
      \mathcal{U} & \text{otherwise.}
    \end{cases}    
    \label{eq:charU}
\end{equation}
Note that all initial states of $\rho_{\mathrm{p}}$ (and therefore $\vartheta$) treated here are symmetric in both real and rapidity space, whereby the solutions obtained using the characteristics above are identical to the ones obtained using periodic boundary conditions. We have checked this numerically.

Next, to account for collisions leading to transverse excitations, we update the quasi-particle distribution following $\rho_{\mathrm{p}} (t + \Delta t) \to \rho_{\mathrm{p}} (t + \Delta t) + 0.5 \Delta t \left( 3 \mathcal{I}[\rho_{\mathrm{p}} (t + \Delta t)] - \mathcal{I}[\rho_{\mathrm{p}} (t)] \right)$.
For the simulations presented in the main text, we assume no atom losses and no initial transverse excitations.
However, we have also performed a number of quasi-1D simulations starting with a thermal occupation of the transverse states, yielding similar results (see Fig.~\ref{fig:measurements}). To estimate the thermal occupation, we fit the initial measured density profiles to the combined density profile of three thermal states $\{ \rho_{\mathrm{p}}^{(l)} \}_{l = 0, 1, 2}$ corresponding to the three lowest transverse levels, with the chemical potential of each thermal state offset by the transverse potential energy $l \hbar \omega_\perp$. The fraction of atoms in each transverse state follow from $\nu_l = \int \mathrm{d}x \, \int \mathrm{d}\theta \, \rho_{\mathrm{p}}^{(l)} / \sum_{j} \int \mathrm{d}x \, \int \mathrm{d}\theta \, \rho_{\mathrm{p}}^{(j)}$.

\section{Apparent relaxation in the Tomonaga Luttinger liquid model} \label{app:LuttingerLiquid}
As argued in the main text, the Tomonaga Luttinger liquid (TLL) model is unsuitable to describe the many-body dynamics observed in our experiments.
Indeed, for the quenches performed, the model does not predict any relaxation of the excited dynamics.
However, due to the statistical nature of our experiment, the model does facilitate an apparent relaxation: Following variations of the atom number, and hence of the speed of sound, between individual experimental repetitions, the dynamics of said repetitions can dephase with respect to one another. 
Thus, when averaging the measured densities, an apparent relaxation of the mean density $n(z,t)$ may occur.

The TLL Hamiltonian can be derived from a perturbative expansion of the Lieb-Liniger Hamiltonian, under the assumption of small density fluctuations and long-wavelength phase-fluctuations \cite{PhysRevA.67.053615}:
\begin{equation}
    H_{TLL} = \int dz\Big[\frac{g}{2}\delta\hat{n}^2(z) + \frac{\hbar^2}{2m}n_0(z)\partial_z\hat{\phi}(z)^2\Big],
    \label{eq:LL_hamiltonian}
\end{equation}
where $\delta\hat{n}$ are the density fluctuations relative to the background $n_0$, and $\hat{\phi}$ is the phase of the quasi-condensate.
In the box trap, the eigenfunctions of the TLL Hamiltonian are non-interacting phononic modes of frequencies $\omega_k = v_{\mathrm{s}}k$, where $v_{\mathrm{s}} = \sqrt{g n_0/m}$ and $k = \frac{2\pi}{L}l$, with $l=1,2,3,...$ ~\cite{RauerThesis}.
In the limit of low-energy excitations, that means low temperatures and weak perturbations, the geometric quench implemented in our experiment can be viewed as the excitation of the coherent population of a single phononic eigenmode.

\begin{figure}
\center
\includegraphics[width = 0.9\columnwidth]{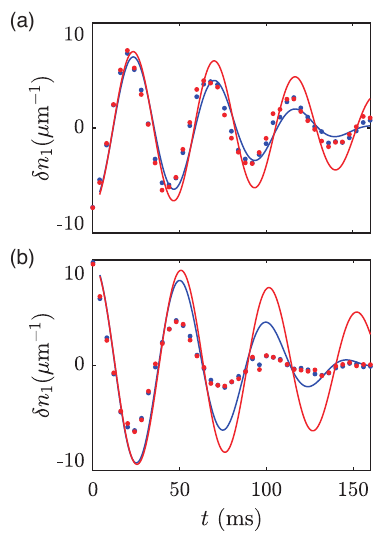}
\caption{\label{fig:LL} Damping of the excited mode given two different atom number distributions with variance $\Delta N = 20\%$ (in blue) and $\Delta N = 10\%$ (in red). The dotted-lines correspond to the damping extracted from experimental data, while the continuous lines show the TLL apparent relaxation expected assuming only dephasing among different realizations. The results shown refer to the two low-amplitude quenches illustrated in Fig.~\ref{fig:carpets_and_modes}: the damping in (\textbf{a}) corresponds to the coldest quench, whose temperature is $T = 46$nK, and the damping in (\textbf{b}) corresponds to the hotter one, with $T = 120$nK.}
\end{figure}

In the following, we will demonstrate that even upon considering such statistical sources of relaxation, low-energy field theories remain unable to capture the observed dynamics of the experiment. 
To this end we consider the two quenches of Fig.~\ref{fig:carpets_and_modes} with low initial mode amplitude and temperatures of $T = 46$nK and $T = 120$nK.
To study how a given variation in the atom number affects the dynamics of the mean density, we post-select the full data set of measurements to obtain subsets with variations of $\Delta N = 20\%$ and $\Delta N = 10\%$. 
Note that following post-selection we ensure that the atom numbers of the subsets remains normally distributed around the same mean value.
From the selected subset we compute the expectation value of the density perturbation $\delta n(z,t)$ and extract the evolution of the addressed mode $\delta n_1 (t)$, as described in the main text.
The resulting evolution of $\delta n_1 (t)$ is plotted in Fig.~\ref{fig:LL} for the $\Delta N = 20\%$ and $\Delta N = 10\%$ subsets as blue and red dots, respectively.
As evident from the figure, the observed relaxation exhibits no change following the post-selection.
Further, we confirm no significant increase in the error of the mean.
For comparison, we compute the expected relaxation of dynamics following the Tomonaga Luttinger liquid model for the two subsets.
Here we model each realization in the subsets as the excitation of a single phononic mode progataed according to Eq.~\eqref{eq:LL_hamiltonian}.
Averaging over the different realizations produces the results plotted in Fig.~\ref{fig:LL} as continuous lines.
As one can see, the TLL apparent relaxation strongly depends on the variance of the atom number; the greater the variation of the speed of sound, the faster the damping.
Such behaviour is not reflected in the measured dynamics, again demonstrating the inability of the TLL model in describing our experimental system.

\newpage
\bibliography{references}% Produces the bibliography via BibTeX.

\begin{thebibliography}{68}%
\makeatletter
\providecommand \@ifxundefined [1]{%
 \@ifx{#1\undefined}
}%
\providecommand \@ifnum [1]{%
 \ifnum #1\expandafter \@firstoftwo
 \else \expandafter \@secondoftwo
 \fi
}%
\providecommand \@ifx [1]{%
 \ifx #1\expandafter \@firstoftwo
 \else \expandafter \@secondoftwo
 \fi
}%
\providecommand \natexlab [1]{#1}%
\providecommand \enquote  [1]{``#1''}%
\providecommand \bibnamefont  [1]{#1}%
\providecommand \bibfnamefont [1]{#1}%
\providecommand \citenamefont [1]{#1}%
\providecommand \href@noop [0]{\@secondoftwo}%
\providecommand \href [0]{\begingroup \@sanitize@url \@href}%
\providecommand \@href[1]{\@@startlink{#1}\@@href}%
\providecommand \@@href[1]{\endgroup#1\@@endlink}%
\providecommand \@sanitize@url [0]{\catcode `\\12\catcode `\$12\catcode
  `\&12\catcode `\#12\catcode `\^12\catcode `\_12\catcode `\%12\relax}%
\providecommand \@@startlink[1]{}%
\providecommand \@@endlink[0]{}%
\providecommand \url  [0]{\begingroup\@sanitize@url \@url }%
\providecommand \@url [1]{\endgroup\@href {#1}{\urlprefix }}%
\providecommand \urlprefix  [0]{URL }%
\providecommand \Eprint [0]{\href }%
\providecommand \doibase [0]{https://doi.org/}%
\providecommand \selectlanguage [0]{\@gobble}%
\providecommand \bibinfo  [0]{\@secondoftwo}%
\providecommand \bibfield  [0]{\@secondoftwo}%
\providecommand \translation [1]{[#1]}%
\providecommand \BibitemOpen [0]{}%
\providecommand \bibitemStop [0]{}%
\providecommand \bibitemNoStop [0]{.\EOS\space}%
\providecommand \EOS [0]{\spacefactor3000\relax}%
\providecommand \BibitemShut  [1]{\csname bibitem#1\endcsname}%
\let\auto@bib@innerbib\@empty
%</preamble>
\bibitem [{\citenamefont {Bethe}(1931)}]{Bethe1931}%
  \BibitemOpen
  \bibfield  {author} {\bibinfo {author} {\bibfnamefont {H.}~\bibnamefont
  {Bethe}},\ }\bibfield  {title} {\bibinfo {title} {Zur {T}heorie der
  {M}etalle},\ }\href {https://doi.org/10.1007/BF01341708} {\bibfield
  {journal} {\bibinfo  {journal} {Zeitschrift f{\"u}r Physik}\ }\textbf
  {\bibinfo {volume} {71}},\ \bibinfo {pages} {205} (\bibinfo {year}
  {1931})}\BibitemShut {NoStop}%
\bibitem [{\citenamefont {Sutherland}(1998)}]{PhysRevLett.80.3678}%
  \BibitemOpen
  \bibfield  {author} {\bibinfo {author} {\bibfnamefont {B.}~\bibnamefont
  {Sutherland}},\ }\bibfield  {title} {\bibinfo {title} {Exact coherent states
  of a one-dimensional quantum fluid in a time-dependent trapping potential},\
  }\href {https://doi.org/10.1103/PhysRevLett.80.3678} {\bibfield  {journal}
  {\bibinfo  {journal} {Phys. Rev. Lett.}\ }\textbf {\bibinfo {volume} {80}},\
  \bibinfo {pages} {3678} (\bibinfo {year} {1998})}\BibitemShut {NoStop}%
\bibitem [{\citenamefont {Campbell}\ \emph {et~al.}(2015)\citenamefont
  {Campbell}, \citenamefont {Gangardt},\ and\ \citenamefont
  {Kheruntsyan}}]{PhysRevLett.114.125302}%
  \BibitemOpen
  \bibfield  {author} {\bibinfo {author} {\bibfnamefont {A.~S.}\ \bibnamefont
  {Campbell}}, \bibinfo {author} {\bibfnamefont {D.~M.}\ \bibnamefont
  {Gangardt}},\ and\ \bibinfo {author} {\bibfnamefont {K.~V.}\ \bibnamefont
  {Kheruntsyan}},\ }\bibfield  {title} {\bibinfo {title} {Sudden expansion of a
  one-dimensional {B}ose gas from power-law traps},\ }\href
  {https://doi.org/10.1103/PhysRevLett.114.125302} {\bibfield  {journal}
  {\bibinfo  {journal} {Phys. Rev. Lett.}\ }\textbf {\bibinfo {volume} {114}},\
  \bibinfo {pages} {125302} (\bibinfo {year} {2015})}\BibitemShut {NoStop}%
\bibitem [{\citenamefont {Lieb}\ and\ \citenamefont
  {Liniger}(1963)}]{lieb1963exact}%
  \BibitemOpen
  \bibfield  {author} {\bibinfo {author} {\bibfnamefont {E.~H.}\ \bibnamefont
  {Lieb}}\ and\ \bibinfo {author} {\bibfnamefont {W.}~\bibnamefont {Liniger}},\
  }\bibfield  {title} {\bibinfo {title} {Exact analysis of an interacting
  {B}ose gas. {I}. {T}he general solution and the ground state},\ }\href
  {https://doi.org/10.1103/PhysRev.130.1605} {\bibfield  {journal} {\bibinfo
  {journal} {Phys. Rev.}\ }\textbf {\bibinfo {volume} {130}},\ \bibinfo {pages}
  {1605} (\bibinfo {year} {1963})}\BibitemShut {NoStop}%
\bibitem [{\citenamefont {Yang}\ and\ \citenamefont
  {Yang}(1969)}]{doi:10.1063/1.1664947}%
  \BibitemOpen
  \bibfield  {author} {\bibinfo {author} {\bibfnamefont {C.~N.}\ \bibnamefont
  {Yang}}\ and\ \bibinfo {author} {\bibfnamefont {C.~P.}\ \bibnamefont
  {Yang}},\ }\bibfield  {title} {\bibinfo {title} {Thermodynamics of a
  one‐dimensional system of bosons with repulsive delta‐function
  interaction},\ }\href {https://doi.org/10.1063/1.1664947} {\bibfield
  {journal} {\bibinfo  {journal} {J. Math. Phys.}\ }\textbf {\bibinfo {volume}
  {10}},\ \bibinfo {pages} {1115} (\bibinfo {year} {1969})}\BibitemShut
  {NoStop}%
\bibitem [{\citenamefont {Cowell}\ \emph {et~al.}(2002)\citenamefont {Cowell},
  \citenamefont {Heiselberg}, \citenamefont {Mazets}, \citenamefont {Morales},
  \citenamefont {Pandharipande},\ and\ \citenamefont
  {Pethick}}]{PhysRevLett.88.210403}%
  \BibitemOpen
  \bibfield  {author} {\bibinfo {author} {\bibfnamefont {S.}~\bibnamefont
  {Cowell}}, \bibinfo {author} {\bibfnamefont {H.}~\bibnamefont {Heiselberg}},
  \bibinfo {author} {\bibfnamefont {I.~E.}\ \bibnamefont {Mazets}}, \bibinfo
  {author} {\bibfnamefont {J.}~\bibnamefont {Morales}}, \bibinfo {author}
  {\bibfnamefont {V.~R.}\ \bibnamefont {Pandharipande}},\ and\ \bibinfo
  {author} {\bibfnamefont {C.~J.}\ \bibnamefont {Pethick}},\ }\bibfield
  {title} {\bibinfo {title} {Cold {B}ose gases with large scattering lengths},\
  }\href {https://doi.org/10.1103/PhysRevLett.88.210403} {\bibfield  {journal}
  {\bibinfo  {journal} {Phys. Rev. Lett.}\ }\textbf {\bibinfo {volume} {88}},\
  \bibinfo {pages} {210403} (\bibinfo {year} {2002})}\BibitemShut {NoStop}%
\bibitem [{\citenamefont {Rigol}\ \emph {et~al.}(2008)\citenamefont {Rigol},
  \citenamefont {Dunjko},\ and\ \citenamefont {Olshanii}}]{Rigol2008}%
  \BibitemOpen
  \bibfield  {author} {\bibinfo {author} {\bibfnamefont {M.}~\bibnamefont
  {Rigol}}, \bibinfo {author} {\bibfnamefont {V.}~\bibnamefont {Dunjko}},\ and\
  \bibinfo {author} {\bibfnamefont {M.}~\bibnamefont {Olshanii}},\ }\bibfield
  {title} {\bibinfo {title} {{Thermalization and its mechanism for generic
  isolated quantum systems}},\ }\href {https://doi.org/10.1038/nature06838}
  {\bibfield  {journal} {\bibinfo  {journal} {Nature}\ }\textbf {\bibinfo
  {volume} {452}},\ \bibinfo {pages} {854} (\bibinfo {year}
  {2008})}\BibitemShut {NoStop}%
\bibitem [{\citenamefont {Gogolin}\ and\ \citenamefont
  {Eisert}(2016)}]{Gogolin2016}%
  \BibitemOpen
  \bibfield  {author} {\bibinfo {author} {\bibfnamefont {C.}~\bibnamefont
  {Gogolin}}\ and\ \bibinfo {author} {\bibfnamefont {J.}~\bibnamefont
  {Eisert}},\ }\bibfield  {title} {\bibinfo {title} {Equilibration,
  thermalisation, and the emergence of statistical mechanics in closed quantum
  systems},\ }\href {https://doi.org/10.1088/0034-4885/79/5/056001} {\bibfield
  {journal} {\bibinfo  {journal} {Rep. Prog. Phys.}\ }\textbf {\bibinfo
  {volume} {79}},\ \bibinfo {pages} {056001} (\bibinfo {year}
  {2016})}\BibitemShut {NoStop}%
\bibitem [{\citenamefont {Gring}\ \emph {et~al.}(2012)\citenamefont {Gring},
  \citenamefont {Kuhnert}, \citenamefont {Langen}, \citenamefont {Kitagawa},
  \citenamefont {Rauer}, \citenamefont {Schreitl}, \citenamefont {Mazets},
  \citenamefont {Smith}, \citenamefont {Demler},\ and\ \citenamefont
  {Schmiedmayer}}]{Gring1318}%
  \BibitemOpen
  \bibfield  {author} {\bibinfo {author} {\bibfnamefont {M.}~\bibnamefont
  {Gring}}, \bibinfo {author} {\bibfnamefont {M.}~\bibnamefont {Kuhnert}},
  \bibinfo {author} {\bibfnamefont {T.}~\bibnamefont {Langen}}, \bibinfo
  {author} {\bibfnamefont {T.}~\bibnamefont {Kitagawa}}, \bibinfo {author}
  {\bibfnamefont {B.}~\bibnamefont {Rauer}}, \bibinfo {author} {\bibfnamefont
  {M.}~\bibnamefont {Schreitl}}, \bibinfo {author} {\bibfnamefont
  {I.}~\bibnamefont {Mazets}}, \bibinfo {author} {\bibfnamefont {D.~A.}\
  \bibnamefont {Smith}}, \bibinfo {author} {\bibfnamefont {E.}~\bibnamefont
  {Demler}},\ and\ \bibinfo {author} {\bibfnamefont {J.}~\bibnamefont
  {Schmiedmayer}},\ }\bibfield  {title} {\bibinfo {title} {Relaxation and
  prethermalization in an isolated quantum system},\ }\href
  {https://doi.org/10.1126/science.1224953} {\bibfield  {journal} {\bibinfo
  {journal} {Science}\ }\textbf {\bibinfo {volume} {337}},\ \bibinfo {pages}
  {1318} (\bibinfo {year} {2012})}\BibitemShut {NoStop}%
\bibitem [{\citenamefont {Bloch}\ \emph {et~al.}(2008)\citenamefont {Bloch},
  \citenamefont {Dalibard},\ and\ \citenamefont {Zwerger}}]{RevModPhys.80.885}%
  \BibitemOpen
  \bibfield  {author} {\bibinfo {author} {\bibfnamefont {I.}~\bibnamefont
  {Bloch}}, \bibinfo {author} {\bibfnamefont {J.}~\bibnamefont {Dalibard}},\
  and\ \bibinfo {author} {\bibfnamefont {W.}~\bibnamefont {Zwerger}},\
  }\bibfield  {title} {\bibinfo {title} {Many-body physics with ultracold
  gases},\ }\href {https://doi.org/10.1103/RevModPhys.80.885} {\bibfield
  {journal} {\bibinfo  {journal} {Rev. Mod. Phys.}\ }\textbf {\bibinfo {volume}
  {80}},\ \bibinfo {pages} {885} (\bibinfo {year} {2008})}\BibitemShut
  {NoStop}%
\bibitem [{\citenamefont {G\"orlitz}\ \emph {et~al.}(2001)\citenamefont
  {G\"orlitz}, \citenamefont {Vogels}, \citenamefont {Leanhardt}, \citenamefont
  {Raman}, \citenamefont {Gustavson}, \citenamefont {Abo-Shaeer}, \citenamefont
  {Chikkatur}, \citenamefont {Gupta}, \citenamefont {Inouye}, \citenamefont
  {Rosenband},\ and\ \citenamefont {Ketterle}}]{PhysRevLett.87.130402}%
  \BibitemOpen
  \bibfield  {author} {\bibinfo {author} {\bibfnamefont {A.}~\bibnamefont
  {G\"orlitz}}, \bibinfo {author} {\bibfnamefont {J.~M.}\ \bibnamefont
  {Vogels}}, \bibinfo {author} {\bibfnamefont {A.~E.}\ \bibnamefont
  {Leanhardt}}, \bibinfo {author} {\bibfnamefont {C.}~\bibnamefont {Raman}},
  \bibinfo {author} {\bibfnamefont {T.~L.}\ \bibnamefont {Gustavson}}, \bibinfo
  {author} {\bibfnamefont {J.~R.}\ \bibnamefont {Abo-Shaeer}}, \bibinfo
  {author} {\bibfnamefont {A.~P.}\ \bibnamefont {Chikkatur}}, \bibinfo {author}
  {\bibfnamefont {S.}~\bibnamefont {Gupta}}, \bibinfo {author} {\bibfnamefont
  {S.}~\bibnamefont {Inouye}}, \bibinfo {author} {\bibfnamefont
  {T.}~\bibnamefont {Rosenband}},\ and\ \bibinfo {author} {\bibfnamefont
  {W.}~\bibnamefont {Ketterle}},\ }\bibfield  {title} {\bibinfo {title}
  {Realization of {B}ose-{E}instein condensates in lower dimensions},\ }\href
  {https://doi.org/10.1103/PhysRevLett.87.130402} {\bibfield  {journal}
  {\bibinfo  {journal} {\textit{Phys. Rev. Lett.}}\ }\textbf {\bibinfo {volume}
  {87}},\ \bibinfo {pages} {130402} (\bibinfo {year} {2001})}\BibitemShut
  {NoStop}%
\bibitem [{\citenamefont {Greiner}\ \emph {et~al.}(2001)\citenamefont
  {Greiner}, \citenamefont {Bloch}, \citenamefont {Mandel}, \citenamefont
  {H\"ansch},\ and\ \citenamefont {Esslinger}}]{PhysRevLett.87.160405}%
  \BibitemOpen
  \bibfield  {author} {\bibinfo {author} {\bibfnamefont {M.}~\bibnamefont
  {Greiner}}, \bibinfo {author} {\bibfnamefont {I.}~\bibnamefont {Bloch}},
  \bibinfo {author} {\bibfnamefont {O.}~\bibnamefont {Mandel}}, \bibinfo
  {author} {\bibfnamefont {T.~W.}\ \bibnamefont {H\"ansch}},\ and\ \bibinfo
  {author} {\bibfnamefont {T.}~\bibnamefont {Esslinger}},\ }\bibfield  {title}
  {\bibinfo {title} {Exploring phase coherence in a 2{D} lattice of
  {B}ose-{E}instein condensates},\ }\href
  {https://doi.org/10.1103/PhysRevLett.87.160405} {\bibfield  {journal}
  {\bibinfo  {journal} {\textit{Phys. Rev. Lett.}}\ }\textbf {\bibinfo {volume}
  {87}},\ \bibinfo {pages} {160405} (\bibinfo {year} {2001})}\BibitemShut
  {NoStop}%
\bibitem [{\citenamefont {Kinoshita}\ \emph {et~al.}(2004)\citenamefont
  {Kinoshita}, \citenamefont {Wenger},\ and\ \citenamefont
  {Weiss}}]{doi:10.1126/science.1100700}%
  \BibitemOpen
  \bibfield  {author} {\bibinfo {author} {\bibfnamefont {T.}~\bibnamefont
  {Kinoshita}}, \bibinfo {author} {\bibfnamefont {T.}~\bibnamefont {Wenger}},\
  and\ \bibinfo {author} {\bibfnamefont {D.~S.}\ \bibnamefont {Weiss}},\
  }\bibfield  {title} {\bibinfo {title} {Observation of a one-dimensional
  {T}onks-{G}irardeau gas},\ }\href {https://doi.org/10.1126/science.1100700}
  {\bibfield  {journal} {\bibinfo  {journal} {Science}\ }\textbf {\bibinfo
  {volume} {305}},\ \bibinfo {pages} {1125} (\bibinfo {year}
  {2004})}\BibitemShut {NoStop}%
\bibitem [{\citenamefont {Kr\"uger}\ \emph {et~al.}(2010)\citenamefont
  {Kr\"uger}, \citenamefont {Hofferberth}, \citenamefont {Mazets},
  \citenamefont {Lesanovsky},\ and\ \citenamefont
  {Schmiedmayer}}]{PhysRevLett.105.265302}%
  \BibitemOpen
  \bibfield  {author} {\bibinfo {author} {\bibfnamefont {P.}~\bibnamefont
  {Kr\"uger}}, \bibinfo {author} {\bibfnamefont {S.}~\bibnamefont
  {Hofferberth}}, \bibinfo {author} {\bibfnamefont {I.~E.}\ \bibnamefont
  {Mazets}}, \bibinfo {author} {\bibfnamefont {I.}~\bibnamefont {Lesanovsky}},\
  and\ \bibinfo {author} {\bibfnamefont {J.}~\bibnamefont {Schmiedmayer}},\
  }\bibfield  {title} {\bibinfo {title} {Weakly interacting {B}ose gas in the
  one-dimensional limit},\ }\href
  {https://doi.org/10.1103/PhysRevLett.105.265302} {\bibfield  {journal}
  {\bibinfo  {journal} {Phys. Rev. Lett.}\ }\textbf {\bibinfo {volume} {105}},\
  \bibinfo {pages} {265302} (\bibinfo {year} {2010})}\BibitemShut {NoStop}%
\bibitem [{\citenamefont {Gerbier}(2004)}]{Gerbier_2004}%
  \BibitemOpen
  \bibfield  {author} {\bibinfo {author} {\bibfnamefont {F.}~\bibnamefont
  {Gerbier}},\ }\bibfield  {title} {\bibinfo {title} {Quasi-1{D}
  {B}ose-{E}instein condensates in the dimensional crossover regime},\ }\href
  {https://doi.org/10.1209/epl/i2004-10035-7} {\bibfield  {journal} {\bibinfo
  {journal} {\textit{EPL}}\ }\textbf {\bibinfo {volume} {66}},\ \bibinfo
  {pages} {771} (\bibinfo {year} {2004})}\BibitemShut {NoStop}%
\bibitem [{\citenamefont {Salasnich}\ \emph {et~al.}(2002)\citenamefont
  {Salasnich}, \citenamefont {Parola},\ and\ \citenamefont
  {Reatto}}]{PhysRevA.65.043614}%
  \BibitemOpen
  \bibfield  {author} {\bibinfo {author} {\bibfnamefont {L.}~\bibnamefont
  {Salasnich}}, \bibinfo {author} {\bibfnamefont {A.}~\bibnamefont {Parola}},\
  and\ \bibinfo {author} {\bibfnamefont {L.}~\bibnamefont {Reatto}},\
  }\bibfield  {title} {\bibinfo {title} {Effective wave equations for the
  dynamics of cigar-shaped and disk-shaped bose condensates},\ }\href
  {https://doi.org/10.1103/PhysRevA.65.043614} {\bibfield  {journal} {\bibinfo
  {journal} {Phys. Rev. A}\ }\textbf {\bibinfo {volume} {65}},\ \bibinfo
  {pages} {043614} (\bibinfo {year} {2002})}\BibitemShut {NoStop}%
\bibitem [{\citenamefont {Li}\ \emph {et~al.}(2020)\citenamefont {Li},
  \citenamefont {Zhou}, \citenamefont {Mazets}, \citenamefont {Stimming},
  \citenamefont {M{\o}ller}, \citenamefont {Zhu}, \citenamefont {Zhai},
  \citenamefont {Xiong}, \citenamefont {Zhou}, \citenamefont {Chen},\ and\
  \citenamefont {Schmiedmayer}}]{10.21468/SciPostPhys.9.4.058}%
  \BibitemOpen
  \bibfield  {author} {\bibinfo {author} {\bibfnamefont {C.}~\bibnamefont
  {Li}}, \bibinfo {author} {\bibfnamefont {T.}~\bibnamefont {Zhou}}, \bibinfo
  {author} {\bibfnamefont {I.}~\bibnamefont {Mazets}}, \bibinfo {author}
  {\bibfnamefont {H.-P.}\ \bibnamefont {Stimming}}, \bibinfo {author}
  {\bibfnamefont {F.~S.}\ \bibnamefont {M{\o}ller}}, \bibinfo {author}
  {\bibfnamefont {Z.}~\bibnamefont {Zhu}}, \bibinfo {author} {\bibfnamefont
  {Y.}~\bibnamefont {Zhai}}, \bibinfo {author} {\bibfnamefont {W.}~\bibnamefont
  {Xiong}}, \bibinfo {author} {\bibfnamefont {X.}~\bibnamefont {Zhou}},
  \bibinfo {author} {\bibfnamefont {X.}~\bibnamefont {Chen}},\ and\ \bibinfo
  {author} {\bibfnamefont {J.}~\bibnamefont {Schmiedmayer}},\ }\bibfield
  {title} {\bibinfo {title} {{Relaxation of Bosons in One Dimension and the
  Onset of Dimensional Crossover}},\ }\href
  {https://doi.org/10.21468/SciPostPhys.9.4.058} {\bibfield  {journal}
  {\bibinfo  {journal} {SciPost Phys.}\ }\textbf {\bibinfo {volume} {9}},\
  \bibinfo {pages} {58} (\bibinfo {year} {2020})}\BibitemShut {NoStop}%
\bibitem [{\citenamefont {Durnin}\ \emph {et~al.}(2021)\citenamefont {Durnin},
  \citenamefont {Bhaseen},\ and\ \citenamefont
  {Doyon}}]{PhysRevLett.127.130601}%
  \BibitemOpen
  \bibfield  {author} {\bibinfo {author} {\bibfnamefont {J.}~\bibnamefont
  {Durnin}}, \bibinfo {author} {\bibfnamefont {M.~J.}\ \bibnamefont
  {Bhaseen}},\ and\ \bibinfo {author} {\bibfnamefont {B.}~\bibnamefont
  {Doyon}},\ }\bibfield  {title} {\bibinfo {title} {Nonequilibrium dynamics and
  weakly broken integrability},\ }\href
  {https://doi.org/10.1103/PhysRevLett.127.130601} {\bibfield  {journal}
  {\bibinfo  {journal} {Phys. Rev. Lett.}\ }\textbf {\bibinfo {volume} {127}},\
  \bibinfo {pages} {130601} (\bibinfo {year} {2021})}\BibitemShut {NoStop}%
\bibitem [{\citenamefont {Bastianello}\ \emph {et~al.}(2021)\citenamefont
  {Bastianello}, \citenamefont {Luca},\ and\ \citenamefont
  {Vasseur}}]{bastianello2021hydrodynamics}%
  \BibitemOpen
  \bibfield  {author} {\bibinfo {author} {\bibfnamefont {A.}~\bibnamefont
  {Bastianello}}, \bibinfo {author} {\bibfnamefont {A.~D.}\ \bibnamefont
  {Luca}},\ and\ \bibinfo {author} {\bibfnamefont {R.}~\bibnamefont
  {Vasseur}},\ }\bibfield  {title} {\bibinfo {title} {Hydrodynamics of weak
  integrability breaking},\ }\href {https://doi.org/10.1088/1742-5468/ac26b2}
  {\bibfield  {journal} {\bibinfo  {journal} {J. Stat. Mech.}\ }\textbf
  {\bibinfo {volume} {2021}},\ \bibinfo {pages} {114003} (\bibinfo {year}
  {2021})}\BibitemShut {NoStop}%
\bibitem [{\citenamefont {Friedman}\ \emph {et~al.}(2020)\citenamefont
  {Friedman}, \citenamefont {Gopalakrishnan},\ and\ \citenamefont
  {Vasseur}}]{PhysRevB.101.180302}%
  \BibitemOpen
  \bibfield  {author} {\bibinfo {author} {\bibfnamefont {A.~J.}\ \bibnamefont
  {Friedman}}, \bibinfo {author} {\bibfnamefont {S.}~\bibnamefont
  {Gopalakrishnan}},\ and\ \bibinfo {author} {\bibfnamefont {R.}~\bibnamefont
  {Vasseur}},\ }\bibfield  {title} {\bibinfo {title} {Diffusive hydrodynamics
  from integrability breaking},\ }\href
  {https://doi.org/10.1103/PhysRevB.101.180302} {\bibfield  {journal} {\bibinfo
   {journal} {Phys. Rev. B}\ }\textbf {\bibinfo {volume} {101}},\ \bibinfo
  {pages} {180302} (\bibinfo {year} {2020})}\BibitemShut {NoStop}%
\bibitem [{\citenamefont {Mallayya}\ \emph {et~al.}(2019)\citenamefont
  {Mallayya}, \citenamefont {Rigol},\ and\ \citenamefont
  {De~Roeck}}]{PhysRevX.9.021027}%
  \BibitemOpen
  \bibfield  {author} {\bibinfo {author} {\bibfnamefont {K.}~\bibnamefont
  {Mallayya}}, \bibinfo {author} {\bibfnamefont {M.}~\bibnamefont {Rigol}},\
  and\ \bibinfo {author} {\bibfnamefont {W.}~\bibnamefont {De~Roeck}},\
  }\bibfield  {title} {\bibinfo {title} {Prethermalization and thermalization
  in isolated quantum systems},\ }\href
  {https://doi.org/10.1103/PhysRevX.9.021027} {\bibfield  {journal} {\bibinfo
  {journal} {Phys. Rev. X}\ }\textbf {\bibinfo {volume} {9}},\ \bibinfo {pages}
  {021027} (\bibinfo {year} {2019})}\BibitemShut {NoStop}%
\bibitem [{\citenamefont {Armijo}\ \emph {et~al.}(2011)\citenamefont {Armijo},
  \citenamefont {Jacqmin}, \citenamefont {Kheruntsyan},\ and\ \citenamefont
  {Bouchoule}}]{PhysRevA.83.021605}%
  \BibitemOpen
  \bibfield  {author} {\bibinfo {author} {\bibfnamefont {J.}~\bibnamefont
  {Armijo}}, \bibinfo {author} {\bibfnamefont {T.}~\bibnamefont {Jacqmin}},
  \bibinfo {author} {\bibfnamefont {K.}~\bibnamefont {Kheruntsyan}},\ and\
  \bibinfo {author} {\bibfnamefont {I.}~\bibnamefont {Bouchoule}},\ }\bibfield
  {title} {\bibinfo {title} {Mapping out the quasicondensate transition through
  the dimensional crossover from one to three dimensions},\ }\href
  {https://doi.org/10.1103/PhysRevA.83.021605} {\bibfield  {journal} {\bibinfo
  {journal} {Phys. Rev. A}\ }\textbf {\bibinfo {volume} {83}},\ \bibinfo
  {pages} {021605(R)} (\bibinfo {year} {2011})}\BibitemShut {NoStop}%
\bibitem [{\citenamefont {Paredes}\ \emph {et~al.}(2004)\citenamefont
  {Paredes}, \citenamefont {Widera}, \citenamefont {Murg}, \citenamefont
  {Mandel}, \citenamefont {F{\"o}lling}, \citenamefont {Cirac}, \citenamefont
  {Shlyapnikov}, \citenamefont {H{\"a}nsch},\ and\ \citenamefont
  {Bloch}}]{Paredes2004}%
  \BibitemOpen
  \bibfield  {author} {\bibinfo {author} {\bibfnamefont {B.}~\bibnamefont
  {Paredes}}, \bibinfo {author} {\bibfnamefont {A.}~\bibnamefont {Widera}},
  \bibinfo {author} {\bibfnamefont {V.}~\bibnamefont {Murg}}, \bibinfo {author}
  {\bibfnamefont {O.}~\bibnamefont {Mandel}}, \bibinfo {author} {\bibfnamefont
  {S.}~\bibnamefont {F{\"o}lling}}, \bibinfo {author} {\bibfnamefont
  {I.}~\bibnamefont {Cirac}}, \bibinfo {author} {\bibfnamefont {G.~V.}\
  \bibnamefont {Shlyapnikov}}, \bibinfo {author} {\bibfnamefont {T.~W.}\
  \bibnamefont {H{\"a}nsch}},\ and\ \bibinfo {author} {\bibfnamefont
  {I.}~\bibnamefont {Bloch}},\ }\bibfield  {title} {\bibinfo {title}
  {{T}onks--{G}irardeau gas of ultracold atoms in an optical lattice},\ }\href
  {https://doi.org/10.1038/nature02530} {\bibfield  {journal} {\bibinfo
  {journal} {Nature}\ }\textbf {\bibinfo {volume} {429}},\ \bibinfo {pages}
  {277} (\bibinfo {year} {2004})}\BibitemShut {NoStop}%
\bibitem [{\citenamefont {Wilson}\ \emph {et~al.}(2020)\citenamefont {Wilson},
  \citenamefont {Malvania}, \citenamefont {Le}, \citenamefont {Zhang},
  \citenamefont {Rigol},\ and\ \citenamefont {Weiss}}]{Wilson2020}%
  \BibitemOpen
  \bibfield  {author} {\bibinfo {author} {\bibfnamefont {J.~M.}\ \bibnamefont
  {Wilson}}, \bibinfo {author} {\bibfnamefont {N.}~\bibnamefont {Malvania}},
  \bibinfo {author} {\bibfnamefont {Y.}~\bibnamefont {Le}}, \bibinfo {author}
  {\bibfnamefont {Y.}~\bibnamefont {Zhang}}, \bibinfo {author} {\bibfnamefont
  {M.}~\bibnamefont {Rigol}},\ and\ \bibinfo {author} {\bibfnamefont {D.~S.}\
  \bibnamefont {Weiss}},\ }\bibfield  {title} {\bibinfo {title} {{Observation
  of dynamical fermionization}},\ }\href
  {https://doi.org/10.1126/science.aaz0242} {\bibfield  {journal} {\bibinfo
  {journal} {Science}\ }\textbf {\bibinfo {volume} {367}},\ \bibinfo {pages}
  {1461} (\bibinfo {year} {2020})}\BibitemShut {NoStop}%
\bibitem [{\citenamefont {Castro-Alvaredo}\ \emph {et~al.}(2016)\citenamefont
  {Castro-Alvaredo}, \citenamefont {Doyon},\ and\ \citenamefont
  {Yoshimura}}]{castro2016emergent}%
  \BibitemOpen
  \bibfield  {author} {\bibinfo {author} {\bibfnamefont {O.~A.}\ \bibnamefont
  {Castro-Alvaredo}}, \bibinfo {author} {\bibfnamefont {B.}~\bibnamefont
  {Doyon}},\ and\ \bibinfo {author} {\bibfnamefont {T.}~\bibnamefont
  {Yoshimura}},\ }\bibfield  {title} {\bibinfo {title} {Emergent hydrodynamics
  in integrable quantum systems out of equilibrium},\ }\href
  {https://doi.org/10.1103/PhysRevX.6.041065} {\bibfield  {journal} {\bibinfo
  {journal} {\textit{Phys. Rev. X}}\ }\textbf {\bibinfo {volume} {6}},\
  \bibinfo {pages} {041065} (\bibinfo {year} {2016})}\BibitemShut {NoStop}%
\bibitem [{\citenamefont {Bertini}\ \emph {et~al.}(2016)\citenamefont
  {Bertini}, \citenamefont {Collura}, \citenamefont {De~Nardis},\ and\
  \citenamefont {Fagotti}}]{bertini2016transport}%
  \BibitemOpen
  \bibfield  {author} {\bibinfo {author} {\bibfnamefont {B.}~\bibnamefont
  {Bertini}}, \bibinfo {author} {\bibfnamefont {M.}~\bibnamefont {Collura}},
  \bibinfo {author} {\bibfnamefont {J.}~\bibnamefont {De~Nardis}},\ and\
  \bibinfo {author} {\bibfnamefont {M.}~\bibnamefont {Fagotti}},\ }\bibfield
  {title} {\bibinfo {title} {Transport in out-of-equilibrium {XXZ} chains:
  Exact profiles of charges and currents},\ }\href
  {https://doi.org/10.1103/PhysRevLett.117.207201} {\bibfield  {journal}
  {\bibinfo  {journal} {Phys. Rev. Lett.}\ }\textbf {\bibinfo {volume} {117}},\
  \bibinfo {pages} {207201} (\bibinfo {year} {2016})}\BibitemShut {NoStop}%
\bibitem [{\citenamefont {Schemmer}\ \emph {et~al.}(2019)\citenamefont
  {Schemmer}, \citenamefont {Bouchoule}, \citenamefont {Doyon},\ and\
  \citenamefont {Dubail}}]{schemmer2019generalized}%
  \BibitemOpen
  \bibfield  {author} {\bibinfo {author} {\bibfnamefont {M.}~\bibnamefont
  {Schemmer}}, \bibinfo {author} {\bibfnamefont {I.}~\bibnamefont {Bouchoule}},
  \bibinfo {author} {\bibfnamefont {B.}~\bibnamefont {Doyon}},\ and\ \bibinfo
  {author} {\bibfnamefont {J.}~\bibnamefont {Dubail}},\ }\bibfield  {title}
  {\bibinfo {title} {Generalized {H}ydrodynamics on an atom chip},\ }\href
  {https://doi.org/10.1103/PhysRevLett.122.090601} {\bibfield  {journal}
  {\bibinfo  {journal} {Phys. Rev. Lett.}\ }\textbf {\bibinfo {volume} {122}},\
  \bibinfo {pages} {090601} (\bibinfo {year} {2019})}\BibitemShut {NoStop}%
\bibitem [{\citenamefont {Malvania}\ \emph {et~al.}(2021)\citenamefont
  {Malvania}, \citenamefont {Zhang}, \citenamefont {Le}, \citenamefont
  {Dubail}, \citenamefont {Rigol},\ and\ \citenamefont
  {Weiss}}]{malvania2020generalized}%
  \BibitemOpen
  \bibfield  {author} {\bibinfo {author} {\bibfnamefont {N.}~\bibnamefont
  {Malvania}}, \bibinfo {author} {\bibfnamefont {Y.}~\bibnamefont {Zhang}},
  \bibinfo {author} {\bibfnamefont {Y.}~\bibnamefont {Le}}, \bibinfo {author}
  {\bibfnamefont {J.}~\bibnamefont {Dubail}}, \bibinfo {author} {\bibfnamefont
  {M.}~\bibnamefont {Rigol}},\ and\ \bibinfo {author} {\bibfnamefont {D.~S.}\
  \bibnamefont {Weiss}},\ }\bibfield  {title} {\bibinfo {title} {Generalized
  hydrodynamics in strongly interacting 1d {B}ose gases},\ }\href
  {https://doi.org/10.1126/science.abf0147} {\bibfield  {journal} {\bibinfo
  {journal} {Science}\ }\textbf {\bibinfo {volume} {373}},\ \bibinfo {pages}
  {1129} (\bibinfo {year} {2021})}\BibitemShut {NoStop}%
\bibitem [{\citenamefont {M{\o}ller}\ \emph
  {et~al.}(2021{\natexlab{a}})\citenamefont {M{\o}ller}, \citenamefont {Li},
  \citenamefont {Mazets}, \citenamefont {Stimming}, \citenamefont {Zhou},
  \citenamefont {Zhu}, \citenamefont {Chen},\ and\ \citenamefont
  {Schmiedmayer}}]{PhysRevLett.126.090602}%
  \BibitemOpen
  \bibfield  {author} {\bibinfo {author} {\bibfnamefont {F.}~\bibnamefont
  {M{\o}ller}}, \bibinfo {author} {\bibfnamefont {C.}~\bibnamefont {Li}},
  \bibinfo {author} {\bibfnamefont {I.}~\bibnamefont {Mazets}}, \bibinfo
  {author} {\bibfnamefont {H.-P.}\ \bibnamefont {Stimming}}, \bibinfo {author}
  {\bibfnamefont {T.}~\bibnamefont {Zhou}}, \bibinfo {author} {\bibfnamefont
  {Z.}~\bibnamefont {Zhu}}, \bibinfo {author} {\bibfnamefont {X.}~\bibnamefont
  {Chen}},\ and\ \bibinfo {author} {\bibfnamefont {J.}~\bibnamefont
  {Schmiedmayer}},\ }\bibfield  {title} {\bibinfo {title} {Extension of the
  {G}eneralized {H}ydrodynamics to the dimensional crossover regime},\ }\href
  {https://doi.org/10.1103/PhysRevLett.126.090602} {\bibfield  {journal}
  {\bibinfo  {journal} {Phys. Rev. Lett.}\ }\textbf {\bibinfo {volume} {126}},\
  \bibinfo {pages} {090602} (\bibinfo {year} {2021}{\natexlab{a}})}\BibitemShut
  {NoStop}%
\bibitem [{\citenamefont {Wigner}(1955)}]{PhysRev.98.145}%
  \BibitemOpen
  \bibfield  {author} {\bibinfo {author} {\bibfnamefont {E.~P.}\ \bibnamefont
  {Wigner}},\ }\bibfield  {title} {\bibinfo {title} {Lower limit for the energy
  derivative of the scattering phase shift},\ }\href
  {https://doi.org/10.1103/PhysRev.98.145} {\bibfield  {journal} {\bibinfo
  {journal} {Phys. Rev.}\ }\textbf {\bibinfo {volume} {98}},\ \bibinfo {pages}
  {145} (\bibinfo {year} {1955})}\BibitemShut {NoStop}%
\bibitem [{\citenamefont {Bulchandani}\ \emph {et~al.}(2018)\citenamefont
  {Bulchandani}, \citenamefont {Vasseur}, \citenamefont {Karrasch},\ and\
  \citenamefont {Moore}}]{PhysRevB.97.045407}%
  \BibitemOpen
  \bibfield  {author} {\bibinfo {author} {\bibfnamefont {V.~B.}\ \bibnamefont
  {Bulchandani}}, \bibinfo {author} {\bibfnamefont {R.}~\bibnamefont
  {Vasseur}}, \bibinfo {author} {\bibfnamefont {C.}~\bibnamefont {Karrasch}},\
  and\ \bibinfo {author} {\bibfnamefont {J.~E.}\ \bibnamefont {Moore}},\
  }\bibfield  {title} {\bibinfo {title} {Bethe-{B}oltzmann hydrodynamics and
  spin transport in the {XXZ} chain},\ }\href
  {https://doi.org/10.1103/PhysRevB.97.045407} {\bibfield  {journal} {\bibinfo
  {journal} {Phys. Rev. B}\ }\textbf {\bibinfo {volume} {97}},\ \bibinfo
  {pages} {045407} (\bibinfo {year} {2018})}\BibitemShut {NoStop}%
\bibitem [{\citenamefont {Doyon}\ \emph {et~al.}(2017)\citenamefont {Doyon},
  \citenamefont {Dubail}, \citenamefont {Konik},\ and\ \citenamefont
  {Yoshimura}}]{PhysRevLett.119.195301}%
  \BibitemOpen
  \bibfield  {author} {\bibinfo {author} {\bibfnamefont {B.}~\bibnamefont
  {Doyon}}, \bibinfo {author} {\bibfnamefont {J.}~\bibnamefont {Dubail}},
  \bibinfo {author} {\bibfnamefont {R.}~\bibnamefont {Konik}},\ and\ \bibinfo
  {author} {\bibfnamefont {T.}~\bibnamefont {Yoshimura}},\ }\bibfield  {title}
  {\bibinfo {title} {Large-scale description of interacting one-dimensional
  {B}ose gases: {G}eneralized {H}ydrodynamics supersedes conventional
  hydrodynamics},\ }\href {https://doi.org/10.1103/PhysRevLett.119.195301}
  {\bibfield  {journal} {\bibinfo  {journal} {Phys. Rev. Lett.}\ }\textbf
  {\bibinfo {volume} {119}},\ \bibinfo {pages} {195301} (\bibinfo {year}
  {2017})}\BibitemShut {NoStop}%
\bibitem [{\citenamefont {Bouchoule}\ and\ \citenamefont
  {Dubail}(2022)}]{bouchoule2022generalized}%
  \BibitemOpen
  \bibfield  {author} {\bibinfo {author} {\bibfnamefont {I.}~\bibnamefont
  {Bouchoule}}\ and\ \bibinfo {author} {\bibfnamefont {J.}~\bibnamefont
  {Dubail}},\ }\bibfield  {title} {\bibinfo {title} {Generalized hydrodynamics
  in the one-dimensional {B}ose gas: theory and experiments},\ }\href
  {https://doi.org/10.1088/1742-5468/ac3659} {\bibfield  {journal} {\bibinfo
  {journal} {J. Stat. Mech.}\ }\textbf {\bibinfo {volume} {2022}},\ \bibinfo
  {pages} {014003} (\bibinfo {year} {2022})}\BibitemShut {NoStop}%
\bibitem [{\citenamefont {Olshanii}(1998)}]{Olshanii1998}%
  \BibitemOpen
  \bibfield  {author} {\bibinfo {author} {\bibfnamefont {M.}~\bibnamefont
  {Olshanii}},\ }\bibfield  {title} {\bibinfo {title} {Atomic scattering in the
  presence of an external confinement and a gas of impenetrable bosons},\
  }\href {https://doi.org/10.1103/PhysRevLett.81.938} {\bibfield  {journal}
  {\bibinfo  {journal} {Phys. Rev. Lett.}\ }\textbf {\bibinfo {volume} {81}},\
  \bibinfo {pages} {938} (\bibinfo {year} {1998})}\BibitemShut {NoStop}%
\bibitem [{\citenamefont {Sutherland}(1968)}]{Sutherland1968}%
  \BibitemOpen
  \bibfield  {author} {\bibinfo {author} {\bibfnamefont {B.}~\bibnamefont
  {Sutherland}},\ }\bibfield  {title} {\bibinfo {title} {Further results for
  the many-body problem in one dimension},\ }\href
  {https://doi.org/10.1103/PhysRevLett.20.98} {\bibfield  {journal} {\bibinfo
  {journal} {Phys. Rev. Lett.}\ }\textbf {\bibinfo {volume} {20}},\ \bibinfo
  {pages} {98} (\bibinfo {year} {1968})}\BibitemShut {NoStop}%
\bibitem [{\citenamefont {Klauser}\ and\ \citenamefont
  {Caux}(2011)}]{klauser2011}%
  \BibitemOpen
  \bibfield  {author} {\bibinfo {author} {\bibfnamefont {A.}~\bibnamefont
  {Klauser}}\ and\ \bibinfo {author} {\bibfnamefont {J.-S.}\ \bibnamefont
  {Caux}},\ }\bibfield  {title} {\bibinfo {title} {{Equilibrium thermodynamic
  properties of interacting two-component bosons in one dimension}},\ }\href
  {https://doi.org/10.1103/PhysRevA.84.033604} {\bibfield  {journal} {\bibinfo
  {journal} {Phys. Rev. A}\ }\textbf {\bibinfo {volume} {84}},\ \bibinfo
  {pages} {033604} (\bibinfo {year} {2011})}\BibitemShut {NoStop}%
\bibitem [{\citenamefont {Gu}\ \emph {et~al.}(2002)\citenamefont {Gu},
  \citenamefont {Li}, \citenamefont {Ying},\ and\ \citenamefont
  {Zhao}}]{doi:10.1142/S0217979202011895}%
  \BibitemOpen
  \bibfield  {author} {\bibinfo {author} {\bibfnamefont {S.-J.}\ \bibnamefont
  {Gu}}, \bibinfo {author} {\bibfnamefont {Y.-Q.}\ \bibnamefont {Li}}, \bibinfo
  {author} {\bibfnamefont {Z.-J.}\ \bibnamefont {Ying}},\ and\ \bibinfo
  {author} {\bibfnamefont {X.-A.}\ \bibnamefont {Zhao}},\ }\bibfield  {title}
  {\bibinfo {title} {Thermodynamics of two component bosons in one dimension},\
  }\href {https://doi.org/10.1142/S0217979202011895} {\bibfield  {journal}
  {\bibinfo  {journal} {International Journal of Modern Physics B}\ }\textbf
  {\bibinfo {volume} {16}},\ \bibinfo {pages} {2137} (\bibinfo {year}
  {2002})}\BibitemShut {NoStop}%
\bibitem [{\citenamefont {Guan}\ \emph {et~al.}(2007)\citenamefont {Guan},
  \citenamefont {Batchelor},\ and\ \citenamefont
  {Takahashi}}]{PhysRevA.76.043617}%
  \BibitemOpen
  \bibfield  {author} {\bibinfo {author} {\bibfnamefont {X.-W.}\ \bibnamefont
  {Guan}}, \bibinfo {author} {\bibfnamefont {M.~T.}\ \bibnamefont
  {Batchelor}},\ and\ \bibinfo {author} {\bibfnamefont {M.}~\bibnamefont
  {Takahashi}},\ }\bibfield  {title} {\bibinfo {title} {Ferromagnetic behavior
  in the strongly interacting two-component bose gas},\ }\href
  {https://doi.org/10.1103/PhysRevA.76.043617} {\bibfield  {journal} {\bibinfo
  {journal} {Phys. Rev. A}\ }\textbf {\bibinfo {volume} {76}},\ \bibinfo
  {pages} {043617} (\bibinfo {year} {2007})}\BibitemShut {NoStop}%
\bibitem [{\citenamefont {Yang}(1967)}]{PhysRevLett.19.1312}%
  \BibitemOpen
  \bibfield  {author} {\bibinfo {author} {\bibfnamefont {C.~N.}\ \bibnamefont
  {Yang}},\ }\bibfield  {title} {\bibinfo {title} {Some exact results for the
  many-body problem in one dimension with repulsive delta-function
  interaction},\ }\href {https://doi.org/10.1103/PhysRevLett.19.1312}
  {\bibfield  {journal} {\bibinfo  {journal} {Phys. Rev. Lett.}\ }\textbf
  {\bibinfo {volume} {19}},\ \bibinfo {pages} {1312} (\bibinfo {year}
  {1967})}\BibitemShut {NoStop}%
\bibitem [{\citenamefont {Reichel}\ and\ \citenamefont
  {Vuleti{\'c}}(2011)}]{atomchips}%
  \BibitemOpen
  \bibfield  {author} {\bibinfo {author} {\bibfnamefont {J.}~\bibnamefont
  {Reichel}}\ and\ \bibinfo {author} {\bibfnamefont {V.}~\bibnamefont
  {Vuleti{\'c}}},\ }\href {https://doi.org/10.1002/9783527633357} {\emph
  {\bibinfo {title} {{\textit{Atom Chips}}}}}\ (\bibinfo  {publisher}
  {Wiley-VCH},\ \bibinfo {address} {Weinheim, Germany},\ \bibinfo {year}
  {2011})\BibitemShut {NoStop}%
\bibitem [{\citenamefont {Manz}\ \emph {et~al.}(2010)\citenamefont {Manz},
  \citenamefont {B\"ucker}, \citenamefont {Betz}, \citenamefont {Koller},
  \citenamefont {Hofferberth}, \citenamefont {Mazets}, \citenamefont
  {Imambekov}, \citenamefont {Demler}, \citenamefont {Perrin}, \citenamefont
  {Schmiedmayer},\ and\ \citenamefont {Schumm}}]{PhysRevA.81.031610}%
  \BibitemOpen
  \bibfield  {author} {\bibinfo {author} {\bibfnamefont {S.}~\bibnamefont
  {Manz}}, \bibinfo {author} {\bibfnamefont {R.}~\bibnamefont {B\"ucker}},
  \bibinfo {author} {\bibfnamefont {T.}~\bibnamefont {Betz}}, \bibinfo {author}
  {\bibfnamefont {C.}~\bibnamefont {Koller}}, \bibinfo {author} {\bibfnamefont
  {S.}~\bibnamefont {Hofferberth}}, \bibinfo {author} {\bibfnamefont {I.~E.}\
  \bibnamefont {Mazets}}, \bibinfo {author} {\bibfnamefont {A.}~\bibnamefont
  {Imambekov}}, \bibinfo {author} {\bibfnamefont {E.}~\bibnamefont {Demler}},
  \bibinfo {author} {\bibfnamefont {A.}~\bibnamefont {Perrin}}, \bibinfo
  {author} {\bibfnamefont {J.}~\bibnamefont {Schmiedmayer}},\ and\ \bibinfo
  {author} {\bibfnamefont {T.}~\bibnamefont {Schumm}},\ }\bibfield  {title}
  {\bibinfo {title} {Two-point density correlations of quasicondensates in free
  expansion},\ }\href {https://doi.org/10.1103/PhysRevA.81.031610} {\bibfield
  {journal} {\bibinfo  {journal} {Phys. Rev. A}\ }\textbf {\bibinfo {volume}
  {81}},\ \bibinfo {pages} {031610(R)} (\bibinfo {year} {2010})}\BibitemShut
  {NoStop}%
\bibitem [{\citenamefont {M{\o}ller}\ \emph
  {et~al.}(2021{\natexlab{b}})\citenamefont {M{\o}ller}, \citenamefont
  {Schweigler}, \citenamefont {Tajik}, \citenamefont {Sabino}, \citenamefont
  {Cataldini}, \citenamefont {Ji},\ and\ \citenamefont
  {Schmiedmayer}}]{PhysRevA.104.043305}%
  \BibitemOpen
  \bibfield  {author} {\bibinfo {author} {\bibfnamefont {F.}~\bibnamefont
  {M{\o}ller}}, \bibinfo {author} {\bibfnamefont {T.}~\bibnamefont
  {Schweigler}}, \bibinfo {author} {\bibfnamefont {M.}~\bibnamefont {Tajik}},
  \bibinfo {author} {\bibfnamefont {J.}~\bibnamefont {Sabino}}, \bibinfo
  {author} {\bibfnamefont {F.}~\bibnamefont {Cataldini}}, \bibinfo {author}
  {\bibfnamefont {S.-C.}\ \bibnamefont {Ji}},\ and\ \bibinfo {author}
  {\bibfnamefont {J.}~\bibnamefont {Schmiedmayer}},\ }\bibfield  {title}
  {\bibinfo {title} {Thermometry of one-dimensional {B}ose gases with neural
  networks},\ }\href {https://doi.org/10.1103/PhysRevA.104.043305} {\bibfield
  {journal} {\bibinfo  {journal} {Phys. Rev. A}\ }\textbf {\bibinfo {volume}
  {104}},\ \bibinfo {pages} {043305} (\bibinfo {year}
  {2021}{\natexlab{b}})}\BibitemShut {NoStop}%
\bibitem [{\citenamefont
  {Schweigler}(2019)}]{https://doi.org/10.48550/arxiv.1908.00422}%
  \BibitemOpen
  \bibfield  {author} {\bibinfo {author} {\bibfnamefont {T.}~\bibnamefont
  {Schweigler}},\ }\href {https://doi.org/10.48550/ARXIV.1908.00422} {\bibinfo
  {title} {Correlations and dynamics of tunnel-coupled one-dimensional bose
  gases}} (\bibinfo {year} {2019})\BibitemShut {NoStop}%
\bibitem [{\citenamefont {Tajik}\ \emph {et~al.}(2019)\citenamefont {Tajik},
  \citenamefont {Rauer}, \citenamefont {Schweigler}, \citenamefont {Cataldini},
  \citenamefont {Sabino}, \citenamefont {M{\o}ller}, \citenamefont {Ji},
  \citenamefont {Mazets},\ and\ \citenamefont {Schmiedmayer}}]{Tajik:19}%
  \BibitemOpen
  \bibfield  {author} {\bibinfo {author} {\bibfnamefont {M.}~\bibnamefont
  {Tajik}}, \bibinfo {author} {\bibfnamefont {B.}~\bibnamefont {Rauer}},
  \bibinfo {author} {\bibfnamefont {T.}~\bibnamefont {Schweigler}}, \bibinfo
  {author} {\bibfnamefont {F.}~\bibnamefont {Cataldini}}, \bibinfo {author}
  {\bibfnamefont {J.}~\bibnamefont {Sabino}}, \bibinfo {author} {\bibfnamefont
  {F.~S.}\ \bibnamefont {M{\o}ller}}, \bibinfo {author} {\bibfnamefont {S.-C.}\
  \bibnamefont {Ji}}, \bibinfo {author} {\bibfnamefont {I.~E.}\ \bibnamefont
  {Mazets}},\ and\ \bibinfo {author} {\bibfnamefont {J.}~\bibnamefont
  {Schmiedmayer}},\ }\bibfield  {title} {\bibinfo {title} {Designing arbitrary
  one-dimensional potentials on an atom chip},\ }\href
  {https://doi.org/10.1364/OE.27.033474} {\bibfield  {journal} {\bibinfo
  {journal} {Opt. Express}\ }\textbf {\bibinfo {volume} {27}},\ \bibinfo
  {pages} {33474} (\bibinfo {year} {2019})}\BibitemShut {NoStop}%
\bibitem [{Note1()}]{Note1}%
  \BibitemOpen
  \bibinfo {note} {For the two high temperature realizations, the bottom of the
  trap is switched from a cosine to a flat potential. For the low temperature
  realization, the order of trap configurations is reversed. For small quench
  amplitudes the evolution of the density perturbation only differs by a sign
  for the two quench types (see Fig.~\ref {fig:double_quench}). In Fig.~\ref
  {fig:carpets_and_modes} the sign for the low temperature realization is
  flipped for easier comparison.}\BibitemShut {Stop}%
\bibitem [{\citenamefont {Hofferberth}\ \emph {et~al.}(2008)\citenamefont
  {Hofferberth}, \citenamefont {Lesanovsky}, \citenamefont {Schumm},
  \citenamefont {Imambekov}, \citenamefont {Gritsev}, \citenamefont {Demler},\
  and\ \citenamefont {Schmiedmayer}}]{Hofferberth2008}%
  \BibitemOpen
  \bibfield  {author} {\bibinfo {author} {\bibfnamefont {S.}~\bibnamefont
  {Hofferberth}}, \bibinfo {author} {\bibfnamefont {I.}~\bibnamefont
  {Lesanovsky}}, \bibinfo {author} {\bibfnamefont {T.}~\bibnamefont {Schumm}},
  \bibinfo {author} {\bibfnamefont {A.}~\bibnamefont {Imambekov}}, \bibinfo
  {author} {\bibfnamefont {V.}~\bibnamefont {Gritsev}}, \bibinfo {author}
  {\bibfnamefont {E.}~\bibnamefont {Demler}},\ and\ \bibinfo {author}
  {\bibfnamefont {J.}~\bibnamefont {Schmiedmayer}},\ }\bibfield  {title}
  {\bibinfo {title} {Probing quantum and thermal noise in an interacting
  many-body system},\ }\href {https://doi.org/10.1038/nphys941} {\bibfield
  {journal} {\bibinfo  {journal} {Nature Physics}\ }\textbf {\bibinfo {volume}
  {4}},\ \bibinfo {pages} {489} (\bibinfo {year} {2008})}\BibitemShut {NoStop}%
\bibitem [{\citenamefont {M{\o}ller}\ and\ \citenamefont
  {Schmiedmayer}(2020)}]{10.21468/SciPostPhys.8.3.041}%
  \BibitemOpen
  \bibfield  {author} {\bibinfo {author} {\bibfnamefont {F.~S.}\ \bibnamefont
  {M{\o}ller}}\ and\ \bibinfo {author} {\bibfnamefont {J.}~\bibnamefont
  {Schmiedmayer}},\ }\bibfield  {title} {\bibinfo {title} {{Introducing iFluid:
  a numerical framework for solving hydrodynamical equations in integrable
  models}},\ }\href {https://doi.org/10.21468/SciPostPhys.8.3.041} {\bibfield
  {journal} {\bibinfo  {journal} {SciPost Phys.}\ }\textbf {\bibinfo {volume}
  {8}},\ \bibinfo {pages} {41} (\bibinfo {year} {2020})}\BibitemShut {NoStop}%
\bibitem [{Note2()}]{Note2}%
  \BibitemOpen
  \bibinfo {note} {The slightly slower relaxation of the GHD simulations
  compared to the experiment in Fig.~\ref {fig:carpets_modes_4th_single} is
  likely due to an underestimation of the temperature.}\BibitemShut {Stop}%
\bibitem [{\citenamefont {De~Nardis}\ \emph {et~al.}(2018)\citenamefont
  {De~Nardis}, \citenamefont {Bernard},\ and\ \citenamefont
  {Doyon}}]{PhysRevLett.121.160603}%
  \BibitemOpen
  \bibfield  {author} {\bibinfo {author} {\bibfnamefont {J.}~\bibnamefont
  {De~Nardis}}, \bibinfo {author} {\bibfnamefont {D.}~\bibnamefont {Bernard}},\
  and\ \bibinfo {author} {\bibfnamefont {B.}~\bibnamefont {Doyon}},\ }\bibfield
   {title} {\bibinfo {title} {Hydrodynamic diffusion in integrable systems},\
  }\href {https://doi.org/10.1103/PhysRevLett.121.160603} {\bibfield  {journal}
  {\bibinfo  {journal} {Phys. Rev. Lett.}\ }\textbf {\bibinfo {volume} {121}},\
  \bibinfo {pages} {160603} (\bibinfo {year} {2018})}\BibitemShut {NoStop}%
\bibitem [{\citenamefont {Bastianello}\ \emph {et~al.}(2020)\citenamefont
  {Bastianello}, \citenamefont {De~Luca}, \citenamefont {Doyon},\ and\
  \citenamefont {De~Nardis}}]{PhysRevLett.125.240604}%
  \BibitemOpen
  \bibfield  {author} {\bibinfo {author} {\bibfnamefont {A.}~\bibnamefont
  {Bastianello}}, \bibinfo {author} {\bibfnamefont {A.}~\bibnamefont
  {De~Luca}}, \bibinfo {author} {\bibfnamefont {B.}~\bibnamefont {Doyon}},\
  and\ \bibinfo {author} {\bibfnamefont {J.}~\bibnamefont {De~Nardis}},\
  }\bibfield  {title} {\bibinfo {title} {Thermalization of a trapped
  one-dimensional {B}ose gas via diffusion},\ }\href
  {https://doi.org/10.1103/PhysRevLett.125.240604} {\bibfield  {journal}
  {\bibinfo  {journal} {Phys. Rev. Lett.}\ }\textbf {\bibinfo {volume} {125}},\
  \bibinfo {pages} {240604} (\bibinfo {year} {2020})}\BibitemShut {NoStop}%
\bibitem [{\citenamefont {Haldane}(1981{\natexlab{a}})}]{PhysRevLett.47.1840}%
  \BibitemOpen
  \bibfield  {author} {\bibinfo {author} {\bibfnamefont {F.~D.~M.}\
  \bibnamefont {Haldane}},\ }\bibfield  {title} {\bibinfo {title} {Effective
  harmonic-fluid approach to low-energy properties of one-dimensional quantum
  fluids},\ }\href {https://doi.org/10.1103/PhysRevLett.47.1840} {\bibfield
  {journal} {\bibinfo  {journal} {Phys. Rev. Lett.}\ }\textbf {\bibinfo
  {volume} {47}},\ \bibinfo {pages} {1840} (\bibinfo {year}
  {1981}{\natexlab{a}})}\BibitemShut {NoStop}%
\bibitem [{\citenamefont {Haldane}(1981{\natexlab{b}})}]{Haldane_1981}%
  \BibitemOpen
  \bibfield  {author} {\bibinfo {author} {\bibfnamefont {F.~D.~M.}\
  \bibnamefont {Haldane}},\ }\bibfield  {title} {\bibinfo {title}
  {{\textquotesingle}{L}uttinger liquid theory{\textquotesingle} of
  one-dimensional quantum fluids. {I}. {P}roperties of the {L}uttinger model
  and their extension to the general 1d interacting spinless {F}ermi gas},\
  }\href {https://doi.org/10.1088/0022-3719/14/19/010} {\bibfield  {journal}
  {\bibinfo  {journal} {Journal of Physics C: Solid State Physics}\ }\textbf
  {\bibinfo {volume} {14}},\ \bibinfo {pages} {2585} (\bibinfo {year}
  {1981}{\natexlab{b}})}\BibitemShut {NoStop}%
\bibitem [{\citenamefont {Mora}\ and\ \citenamefont
  {Castin}(2003)}]{PhysRevA.67.053615}%
  \BibitemOpen
  \bibfield  {author} {\bibinfo {author} {\bibfnamefont {C.}~\bibnamefont
  {Mora}}\ and\ \bibinfo {author} {\bibfnamefont {Y.}~\bibnamefont {Castin}},\
  }\bibfield  {title} {\bibinfo {title} {Extension of {B}ogoliubov theory to
  quasicondensates},\ }\href {https://doi.org/10.1103/PhysRevA.67.053615}
  {\bibfield  {journal} {\bibinfo  {journal} {Phys. Rev. A}\ }\textbf {\bibinfo
  {volume} {67}},\ \bibinfo {pages} {053615} (\bibinfo {year}
  {2003})}\BibitemShut {NoStop}%
\bibitem [{\citenamefont {Korepin}\ \emph {et~al.}(1993)\citenamefont
  {Korepin}, \citenamefont {Bogoliubov},\ and\ \citenamefont
  {Izergin}}]{korepin_bogoliubov_izergin_1993}%
  \BibitemOpen
  \bibfield  {author} {\bibinfo {author} {\bibfnamefont {V.~E.}\ \bibnamefont
  {Korepin}}, \bibinfo {author} {\bibfnamefont {N.~M.}\ \bibnamefont
  {Bogoliubov}},\ and\ \bibinfo {author} {\bibfnamefont {A.~G.}\ \bibnamefont
  {Izergin}},\ }\href {https://doi.org/10.1017/CBO9780511628832} {\emph
  {\bibinfo {title} {Quantum Inverse Scattering Method and Correlation
  Functions}}},\ Cambridge Monographs on Mathematical Physics\ (\bibinfo
  {publisher} {Cambridge University Press},\ \bibinfo {year}
  {1993})\BibitemShut {NoStop}%
\bibitem [{\citenamefont {Lieb}(1963)}]{LL2}%
  \BibitemOpen
  \bibfield  {author} {\bibinfo {author} {\bibfnamefont {E.~H.}\ \bibnamefont
  {Lieb}},\ }\bibfield  {title} {\bibinfo {title} {Exact analysis of an
  interacting {B}ose gas. {II}. {T}he excitation spectrum},\ }\href
  {https://doi.org/10.1103/PhysRev.130.1616} {\bibfield  {journal} {\bibinfo
  {journal} {Phys. Rev.}\ }\textbf {\bibinfo {volume} {130}},\ \bibinfo {pages}
  {1616} (\bibinfo {year} {1963})}\BibitemShut {NoStop}%
\bibitem [{Note3()}]{Note3}%
  \BibitemOpen
  \bibinfo {note} {Assuming a linearization of the dressing operation around
  the stationary background (here, zeroth mode of the occupation
  function).}\BibitemShut {Stop}%
\bibitem [{\citenamefont {M{\o}ller}\ \emph {et~al.}(2022)\citenamefont
  {M{\o}ller}, \citenamefont {Erne}, \citenamefont {Mauser}, \citenamefont
  {Schmiedmayer},\ and\ \citenamefont {Mazets}}]{Moller2022}%
  \BibitemOpen
  \bibfield  {author} {\bibinfo {author} {\bibfnamefont {F.}~\bibnamefont
  {M{\o}ller}}, \bibinfo {author} {\bibfnamefont {S.}~\bibnamefont {Erne}},
  \bibinfo {author} {\bibfnamefont {N.~J.}\ \bibnamefont {Mauser}}, \bibinfo
  {author} {\bibfnamefont {J.}~\bibnamefont {Schmiedmayer}},\ and\ \bibinfo
  {author} {\bibfnamefont {I.~E.}\ \bibnamefont {Mazets}},\ }\href
  {https://doi.org/10.48550/ARXIV.2205.15871} {\bibinfo {title} {Bridging
  effective field theories and generalized hydrodynamics}} (\bibinfo {year}
  {2022})\BibitemShut {NoStop}%
\bibitem [{\citenamefont {Panfil}\ and\ \citenamefont
  {Pawełczyk}(2019)}]{10.21468/SciPostPhysCore.1.1.002}%
  \BibitemOpen
  \bibfield  {author} {\bibinfo {author} {\bibfnamefont {M.}~\bibnamefont
  {Panfil}}\ and\ \bibinfo {author} {\bibfnamefont {J.}~\bibnamefont
  {Pawełczyk}},\ }\bibfield  {title} {\bibinfo {title} {{Linearized regime of
  the generalized hydrodynamics with diffusion}},\ }\href
  {https://doi.org/10.21468/SciPostPhysCore.1.1.002} {\bibfield  {journal}
  {\bibinfo  {journal} {SciPost Phys. Core}\ }\textbf {\bibinfo {volume} {1}},\
  \bibinfo {pages} {2} (\bibinfo {year} {2019})}\BibitemShut {NoStop}%
\bibitem [{\citenamefont {Kinoshita}\ \emph {et~al.}(2006)\citenamefont
  {Kinoshita}, \citenamefont {Wenger},\ and\ \citenamefont
  {Weiss}}]{kinoshita2006quantum}%
  \BibitemOpen
  \bibfield  {author} {\bibinfo {author} {\bibfnamefont {T.}~\bibnamefont
  {Kinoshita}}, \bibinfo {author} {\bibfnamefont {T.}~\bibnamefont {Wenger}},\
  and\ \bibinfo {author} {\bibfnamefont {D.~S.}\ \bibnamefont {Weiss}},\
  }\bibfield  {title} {\bibinfo {title} {A quantum {N}ewton's cradle},\ }\href
  {https://doi.org/10.1038/nature04693} {\bibfield  {journal} {\bibinfo
  {journal} {\textit{Nature}}\ }\textbf {\bibinfo {volume} {440}},\ \bibinfo
  {pages} {900} (\bibinfo {year} {2006})}\BibitemShut {NoStop}%
\bibitem [{\citenamefont {Caux}\ \emph {et~al.}(2019)\citenamefont {Caux},
  \citenamefont {Doyon}, \citenamefont {Dubail}, \citenamefont {Konik},\ and\
  \citenamefont {Yoshimura}}]{10.21468/SciPostPhys.6.6.070}%
  \BibitemOpen
  \bibfield  {author} {\bibinfo {author} {\bibfnamefont {J.-S.}\ \bibnamefont
  {Caux}}, \bibinfo {author} {\bibfnamefont {B.}~\bibnamefont {Doyon}},
  \bibinfo {author} {\bibfnamefont {J.}~\bibnamefont {Dubail}}, \bibinfo
  {author} {\bibfnamefont {R.}~\bibnamefont {Konik}},\ and\ \bibinfo {author}
  {\bibfnamefont {T.}~\bibnamefont {Yoshimura}},\ }\bibfield  {title} {\bibinfo
  {title} {{Hydrodynamics of the interacting {B}ose gas in the Quantum {N}ewton
  Cradle setup}},\ }\href {https://doi.org/10.21468/SciPostPhys.6.6.070}
  {\bibfield  {journal} {\bibinfo  {journal} {SciPost Phys.}\ }\textbf
  {\bibinfo {volume} {6}},\ \bibinfo {pages} {70} (\bibinfo {year}
  {2019})}\BibitemShut {NoStop}%
\bibitem [{\citenamefont {Rauer}(2019)}]{RauerThesis}%
  \BibitemOpen
  \bibfield  {author} {\bibinfo {author} {\bibfnamefont {B.}~\bibnamefont
  {Rauer}},\ }\href {https://doi.org/https://doi.org/10.1007/978-3-030-18236-6}
  {\emph {\bibinfo {title} {{\textit{Non-equilibrium dynamics beyond dephasing:
  Recurrences and loss induced cooling in one-dimensional {B}ose gases}}}}}\
  (\bibinfo  {publisher} {Springer, Heidelberg},\ \bibinfo {year}
  {2019})\BibitemShut {NoStop}%
\bibitem [{\citenamefont {Grimm}\ \emph {et~al.}(2000)\citenamefont {Grimm},
  \citenamefont {Weidem{\"u}ller},\ and\ \citenamefont
  {Ovchinnikov}}]{GRIMM200095}%
  \BibitemOpen
  \bibfield  {author} {\bibinfo {author} {\bibfnamefont {R.}~\bibnamefont
  {Grimm}}, \bibinfo {author} {\bibfnamefont {M.}~\bibnamefont
  {Weidem{\"u}ller}},\ and\ \bibinfo {author} {\bibfnamefont {Y.~B.}\
  \bibnamefont {Ovchinnikov}},\ }\bibfield  {title} {\bibinfo {title} {Optical
  dipole traps for neutral atoms}\ }(\bibinfo  {publisher} {Academic Press},\
  \bibinfo {year} {2000})\ pp.\ \bibinfo {pages} {95--170}\BibitemShut
  {NoStop}%
\bibitem [{\citenamefont {Dahl}\ and\ \citenamefont
  {Schleich}(2009)}]{PhysRevA.79.024101}%
  \BibitemOpen
  \bibfield  {author} {\bibinfo {author} {\bibfnamefont {J.~P.}\ \bibnamefont
  {Dahl}}\ and\ \bibinfo {author} {\bibfnamefont {W.~P.}\ \bibnamefont
  {Schleich}},\ }\bibfield  {title} {\bibinfo {title} {State operator,
  constants of the motion, and wigner functions: The two-dimensional isotropic
  harmonic oscillator},\ }\href {https://doi.org/10.1103/PhysRevA.79.024101}
  {\bibfield  {journal} {\bibinfo  {journal} {Phys. Rev. A}\ }\textbf {\bibinfo
  {volume} {79}},\ \bibinfo {pages} {024101} (\bibinfo {year}
  {2009})}\BibitemShut {NoStop}%
\bibitem [{\citenamefont {Abramowitz}\ and\ \citenamefont
  {Stegun}(1965)}]{mabramowitz64:handbook}%
  \BibitemOpen
  \bibinfo {editor} {\bibfnamefont {M.}~\bibnamefont {Abramowitz}}\ and\
  \bibinfo {editor} {\bibfnamefont {I.~A.}\ \bibnamefont {Stegun}},\ eds.,\
  \href@noop {} {\emph {\bibinfo {title} {Handbook of Mathematical Functions
  with Formulas, Graphs and Mathematical Tables}}}\ (\bibinfo  {publisher}
  {Dover Publications, Inc.},\ \bibinfo {address} {New York},\ \bibinfo {year}
  {1965})\ Chap.~\bibinfo {chapter} {22}\BibitemShut {NoStop}%
\bibitem [{\citenamefont {Lamacraft}(2013)}]{PhysRevA.87.012707}%
  \BibitemOpen
  \bibfield  {author} {\bibinfo {author} {\bibfnamefont {A.}~\bibnamefont
  {Lamacraft}},\ }\bibfield  {title} {\bibinfo {title} {Diffractive scattering
  of three particles in one dimension: A simple result for weak violations of
  the yang-baxter equation},\ }\href
  {https://doi.org/10.1103/PhysRevA.87.012707} {\bibfield  {journal} {\bibinfo
  {journal} {Phys. Rev. A}\ }\textbf {\bibinfo {volume} {87}},\ \bibinfo
  {pages} {012707} (\bibinfo {year} {2013})}\BibitemShut {NoStop}%
\bibitem [{\citenamefont {Jimbo}(1990)}]{jimbo1990yang}%
  \BibitemOpen
  \bibfield  {author} {\bibinfo {author} {\bibfnamefont {M.}~\bibnamefont
  {Jimbo}},\ }\href@noop {} {\emph {\bibinfo {title} {{Y}ang-{B}axter Equation
  in Integrable Systems}}},\ Advanced series in mathematical physics\ (\bibinfo
   {publisher} {World Scientific},\ \bibinfo {year} {1990})\BibitemShut
  {NoStop}%
\bibitem [{\citenamefont {Bunker}\ and\ \citenamefont
  {Jensen}(1998)}]{Bunker1998}%
  \BibitemOpen
  \bibfield  {author} {\bibinfo {author} {\bibfnamefont {P.}~\bibnamefont
  {Bunker}}\ and\ \bibinfo {author} {\bibfnamefont {P.}~\bibnamefont
  {Jensen}},\ }\href {https://doi.org/10.1063/1.2995676} {\emph {\bibinfo
  {title} {Molecular Symmetry and Spectroscopy}}},\ Vol.~\bibinfo {volume}
  {32}\ (\bibinfo {year} {1998})\ Chap.~\bibinfo {chapter} {9}\BibitemShut
  {NoStop}%
\bibitem [{\citenamefont {Mazets}\ and\ \citenamefont
  {Schmiedmayer}(2010)}]{Mazets_2010}%
  \BibitemOpen
  \bibfield  {author} {\bibinfo {author} {\bibfnamefont {I.~E.}\ \bibnamefont
  {Mazets}}\ and\ \bibinfo {author} {\bibfnamefont {J.}~\bibnamefont
  {Schmiedmayer}},\ }\bibfield  {title} {\bibinfo {title} {Thermalization in a
  quasi-one-dimensional ultracold bosonic gas},\ }\href
  {https://doi.org/10.1088/1367-2630/12/5/055023} {\bibfield  {journal}
  {\bibinfo  {journal} {New Journal of Physics}\ }\textbf {\bibinfo {volume}
  {12}},\ \bibinfo {pages} {055023} (\bibinfo {year} {2010})}\BibitemShut
  {NoStop}%
\end{thebibliography}%

\end{document}